\UseRawInputEncoding
\documentclass[11pt]{article}
\pdfoutput=1
\usepackage{jheppub}
\usepackage[T1]{fontenc}
\usepackage{cancel,tensor,enumitem,fix-cm}
\usepackage{epsfig}
\usepackage{graphicx}
\usepackage[english]{babel}
\usepackage{amsmath}
\usepackage{textcomp}
\usepackage{multirow}
\usepackage{booktabs}
\usepackage{tikz}
\usepackage{overpic}
\usepackage{bm}
\usepackage{physics}
\usepackage[lofdepth,lotdepth]{subfig}

\newcommand{\be}{\begin{equation}}
\newcommand{\ee}{\end{equation}}
\newcommand{\bea}{\begin{eqnarray}}
\newcommand{\eea}{\end{eqnarray}}
\def\({\left(}
\def\){\right)}
\def\[{\left[}
\def\]{\right]}
\def\cO{{\cal O}}

\makeatletter
\newsavebox\myboxA
\newsavebox\myboxB
\newlength\mylenA
\newcommand*\xoverline[2][0.75]{%
    \sbox{\myboxA}{$\m@th#2$}%
    \setbox\myboxB\null
    \ht\myboxB=\ht\myboxA%
    \dp\myboxB=\dp\myboxA%
    \wd\myboxB=#1\wd\myboxA
    \sbox\myboxB{$\m@th\overline{\copy\myboxB}$}
    \setlength\mylenA{\the\wd\myboxA}
    \addtolength\mylenA{-\the\wd\myboxB}%
    \ifdim\wd\myboxB<\wd\myboxA%
       \rlap{\hskip 0.5\mylenA\usebox\myboxB}{\usebox\myboxA}%
    \else
        \hskip -0.5\mylenA\rlap{\usebox\myboxA}{\hskip 0.5\mylenA\usebox\myboxB}%
    \fi}
\makeatother

\subheader{\begin{flushright}
\texttt{BRX-TH-6679}
\end{flushright}}

\title{Quantum bit threads and holographic entanglement}

\author[a]{Cesar A. Ag\'on,}
\author[b,c]{and Juan F. Pedraza}
\affiliation[a]{Instituto Balseiro, Centro At\'omico Bariloche, 8400-S.C. de Bariloche, R\'io Negro, Argentina}
\affiliation[b]{Department of Physics and Astronomy, University College London, London WC1E 6BT, UK}
\affiliation[c]{Martin Fisher School of Physics, Brandeis University, Waltham MA 02453, USA}
\emailAdd{cesar.agon@cab.cnea.gov.ar}
\emailAdd{j.pedraza@ucl.ac.uk}

\abstract{Quantum corrections to holographic entanglement entropy require knowledge of the bulk quantum state. In this paper, we
derive a novel dual prescription for the generalized entropy that allows us to interpret the leading quantum corrections in a geometric way with minimal input from the bulk state.  The equivalence is proven using tools borrowed from convex optimization. The new prescription does not involve bulk surfaces but instead uses a generalized notion of a flow, which allows for possible sources or sinks in the bulk geometry. In its discrete version, our prescription can alternatively be interpreted in terms of a set of Planck-thickness bit threads, which can be either classical or quantum. This interpretation uncovers an aspect of the generalized entropy
that admits a neat information-theoretic description, namely, the fact that the quantum corrections can be cast in terms of entanglement distillation of the bulk state. We also prove some general properties of our prescription, including nesting and a quantum version of the max multiflow theorem. These properties are used to verify that our proposal respects known inequalities that a von Neumann entropy must satisfy, including subadditivity and strong subadditivity, as well as to investigate the fate of the holographic monogamy. Finally, using the Iyer-Wald formalism we show that for cases with a local modular Hamiltonian there is always a canonical solution to the program that exploits the property of bulk locality. Combining with previous results by Swingle and Van Raamsdonk, we show that the consistency of this special solution requires the semi-classical Einstein's equations to hold for any consistent perturbative bulk quantum state.
}

\begin{document}
\maketitle
\flushbottom

\section{Introduction}

\subsection{General motivation}

Recent progress in quantum gravity has revealed a surprising connection between spacetime and quantum information. The sharpest realization of this connection is formulated in the context of the AdS/CFT correspondence, a remarkable duality between a theory of quantum gravity in the `bulk' of an asymptotically Anti-de Sitter space (AdS) and a strongly-coupled Conformal Field Theory (CFT) living in its lower-dimensional boundary \cite{Maldacena:1997re}. An exciting development that ignited this line of research was the proposed Ryu-Takayanagi (RT) formula, relating the area of certain codimension-2 surfaces in the bulk to the entanglement entropy of subsystems in the dual CFT \cite{Ryu:2006bv}. In the bulk theory, the RT formula can be interpreted as a generalization of the Bekenstein-Hawking formula that computes black hole entropy \cite{Bekenstein:1973ur,Hawking:1974sw}. Conversely, in the dual CFT, the entanglement or von Neumann entropy generalizes the notion of thermodynamic entropy for states that are not necessarily thermal. Even though the RT formula had passed various consistency checks and was known to satisfy all known properties of the von Neumann entropy \cite{Headrick:2013zda}, it was not until \cite{Lewkowycz:2013nqa} that a formal proof of the prescription was provided. The RT formula has been further generalized in a number of ways, including to covariant settings \cite{Hubeny:2007xt,Dong:2016hjy}, to the case of higher curvature gravities (finite-coupling corrections) \cite{Dong:2013qoa,Camps:2013zua}, and when $1/N$ quantum corrections are taken into account \cite{Faulkner:2013ana,Engelhardt:2014gca}.

One of the most interesting applications that emerged from the connection between gravity and quantum information is related to the program of bulk reconstruction. Since the RT surfaces probe the bulk geometry, there have been numerous proposals for reconstructing the bulk metric using entanglement entropies in the dual CFT, in various particular contexts \cite{Czech:2012bh,Balasubramanian:2013lsa,Myers:2014jia,Czech:2014wka,Headrick:2014eia,Czech:2014ppa,Czech:2015qta,Faulkner:2018faa,Roy:2018ehv,Espindola:2017jil,Espindola:2018ozt,Balasubramanian:2018uus,Bao:2019bib,Jokela:2020auu,Bao:2020abm}. Thus, at least intuitively, the RT prescription suggests that spacetime emerges from entanglement \cite{VanRaamsdonk:2009ar,VanRaamsdonk:2010pw,Bianchi:2012ev,Maldacena:2013xja,Balasubramanian:2014sra}, an idea that is summarized by the slogan `geometry from entanglement' or, oftentimes, `it from qubit'. A more refined version of this program relates the dynamics of the bulk metric (subject to appropriate boundary conditions) to the rules of governing the entanglement entropy under changes of the CFT state or CFT Hamiltonian \cite{Swingle:2014uza,Caceres:2016xjz,Czech:2016tqr,Faulkner:2017tkh,Dong:2017xht,Haehl:2017sot,Lewkowycz:2018sgn,Rosso:2020zkk}. Here the RT formula enters again as a fundamental input and the previous slogan is then upgraded to `gravitation from entanglement'. The RT prescription has also inspired various developments connecting spacetime to other topics in quantum information, including tensor networks \cite{Swingle:2009bg,Hayden:2016cfa,Bao:2018pvs,Jahn:2021uqr}, error correction \cite{Almheiri:2014lwa,Pastawski:2015qua,Dong:2016eik,Harlow:2016vwg}, quantum computation \cite{Susskind:2014rva,Brown:2015bva,Brown:2015lvg,Caputa:2017yrh,Couch:2016exn} and quantum teleportation \cite{Gao:2016bin,Maldacena:2017axo,Caceres:2018ehr,Brown:2019hmk,Freivogel:2019lej,Freivogel:2019whb}, among others. Collectively then, statements about gravity are then reinterpreted via holography as statements in quantum information theory and viceversa.

Recently, Headrick and Freedman showed that the RT formula admits a dual description in terms of flows, divergenceless norm-bounded vector fields, or equivalently a set of Planck-thickness ``bit threads'' \cite{Freedman:2016zud}.\footnote{These divergenceless flows are Hodge dual to closed forms called calibrations \cite{Harvey:1982xk}, which are likewise useful to describe holographic entanglement \cite{Bakhmatov:2017ihw}.} The new prescription involves maximizing the flux through a region and follows from the max flow-min cut theorem of network theory. Its formal proof involves further elements from convex optimization and convex relaxation, as well as strong duality \cite{Headrick:2017ucz}.
This new prescription has helped uncover aspects of holographic entanglement and related quantities that were previously unknown, and has provided a nice information-theoretic interpretation of various known properties \cite{Cui:2018dyq,Chen:2018ywy,Hubeny:2018bri,Agon:2018lwq,Ghodrati:2019hnn,Kudler-Flam:2019oru,Du:2019emy,Bao:2019wcf,Harper:2019lff,Agon:2019qgh,Du:2019vwh,Harper:2020wad,Agon:2020mvu,Headrick:2020gyq,Lin:2020yzf,Ghodrati:2020vzm,Bao:2020uku,Lin:2021hqs,Pedraza:2021mkh,Pedraza:2021fgp}. Similar to the RT formula, the bit thread formulation of holographic entanglement entropy has been generalized to the covariant settings \cite{Headrick:toappear} and for CFTs dual to higher curvature gravities \cite{Harper:2018sdd}, though, a version that incorporates quantum or $1/N$ corrections has not been worked out. This will be the main motivation of the present work and the central problem that we will try to address. Along the way, we will explore some general properties that this quantum corrected prescription should satisfy and explore aspects of their physical interpretation. As an application we will also explore dynamical aspects of our proposal, and thus make connection with the program of `gravitation from entanglement' discussed above.

\subsection{Setup and organization of the paper}

Let us now give details of the particular problem that we want to address. At leading order in $G_N\sim 1/N^2$, i.e., at $\mathcal{O}(1/G_N)$, the RT formula give us the entanglement entropy of a boundary region $A$ as the area of a minimal codimension-2 surface in the bulk \cite{Ryu:2006bv},
\be
S[A]=\underset{\gamma_A\sim A}{\text{min}}\left[\frac{A(\gamma_A)}{4 G_N}\right]\, ,
\ee
satisfying the homology condition $\gamma_A\sim A$. This formula gets corrected at $\mathcal{O}(G_N^0)$
\be\label{FLM}
S[A] = \underset{\gamma_A\sim A}{\text{min}}\left[\frac{A(\gamma_A)}{4 G_N}\right] + S_{\text{bulk}}[\Sigma_A]\, ,
\ee
which is known as the Faulkner-Lewkowycz-Maldacena (FLM) formula \cite{Faulkner:2013ana}. This is applicable for states with semi-classical gravity duals, i.e., those described by effective quantum field theory in the bulk living on a curved but classical background. The new term, $S_{\text{bulk}}$, represents the von Neumann entropy of the bulk state reduced to the entanglement wedge,
while $\Sigma_A$ denotes an arbitrary Cauchy slice satisfying $\partial \Sigma_A = \gamma_A\cup A$. A more accurate prescription, valid beyond leading order in $G_N$ is given by the Quantum Extremal Surface (QES) formula, proposed by Engelhardt and Wall \cite{Engelhardt:2014gca}. This involves a minimization of the two terms in (\ref{FLM}), area and bulk entropy:
\be\label{QES}
S[A] = \underset{\tilde{\gamma}_A\sim A}{\text{min}}\left[\frac{A(\tilde{\gamma}_A)}{4 G_N} + S_{\text{bulk}}[\tilde{\Sigma}_A]\right],
\ee
Notice that here we have used tildes to distinguish from the quantities that arise from the pure area minimization. Importantly, the bulk states we will consider are semi-classical states that admit a metric expansion of the form\footnote{More generally, fractional orders can arise in this expansion because quantized gravitons have amplitude $\sqrt{G_N}$. However, as done by FLM \cite{Faulkner:2013ana}, here we ignore graviton fluctuations.}
\be
g_{\mu\nu}=g_{\mu\nu}^{(0)}+G_N g_{\mu\nu}^{(1)}+G_N^2g_{\mu\nu}^{(2)}+\mathcal{O}(G_N^3)\,.
\ee
For these states, the first correction to the QES surface is typically suppressed such that $\tilde{\gamma}_A=\gamma_A+\mathcal{O}(G_N^1)$. However, due to the minimality condition, these corrections do not affect the area \cite{Belin:2018juv,Agon:2020fqs} so both formulas give the same result at $\mathcal{O}(G_N^0)$, i.e.,
\be
S[A] = \underset{\tilde{\gamma}_A\sim A}{\text{min}}\left[\frac{A(\tilde{\gamma}_A)}{4 G_N} + S_{\text{bulk}}[\tilde{\Sigma}_A]\right]\approx \underset{\gamma_A\sim A}{\text{min}}\left[\frac{A(\gamma_A)}{4 G_N}\right] + S_{\text{bulk}}[\Sigma_A]\,.
\ee
Technically, however, the two formulas can \emph{differ} sufficiently close to a phase transition, where the corrections due to the bulk entanglement entropy can induce a discontinuous jump between two different classical saddles. In these situations the surfaces differ at order $\mathcal{O}(G_N^0)$, leading to a difference between the two prescriptions of order $\mathcal{O}(G_N^0)$, and the QES formula gives the correct result (up to non-perturbative corrections that we do not consider here).\footnote{We will discuss these situations in detail in section \ref{multiple-ms}.}

It is interesting to ask how the bit thread prescription \cite{Freedman:2016zud} gets corrected when quantum corrections are taken into account. At the leading order, we have that
\be\label{BTpres}
S[A]=\frac{1}{4G_N}\, \max_{v\in \mathcal{F}}\int_{A} v\,,\qquad {\cal F}\equiv\{v\, \vert\, \nabla\cdot v=0,\, |v|\leq 1\}\,.
\ee
and the equivalence with the RT prescription can be proven using convex optimization and strong duality. A natural question is if the FLM or QES formulas described above admit a similar description which can be proven using the same techniques. The main goal of this paper is to provide an answer to this question. A technical point here is that in order to derive the dual description we will need to restrict ourselves to the first order correction, i.e., the FLM formula or the QES equivalent. It is only in this case that the prescription can be formulated as a convex program and, hence, can be dealt with using convex optimization techniques. We will however, be able to derive a nice interpretation for the leading quantum corrections. In particular, we will see that the \emph{quantum bit thread prescription} that we arrive at can be interpreted as a `geometrization' of these corrections, where both the area and the bulk entropy pieces are unified in a single vector field description.

This paper is organized as follows. In section \ref{sec:FlowProgram} we derive and interpret the new flow prescription that incorporates quantum corrections. This section is divided in three parts. In subsection \ref{sec:heuristic} we present an argument for our proposal based on the Jafferis-Lewkowycz-Maldacena-Suh (JLMS) formula \cite{Jafferis:2015del} ---an operator version of FLM. This argument is thus valid for the leading quantum corrections. In subsection \ref{ConvexOpt} we provide a proof of the dual program using tools of convex optimization. Although we start from a version of the QES formula in this proof, we explain the reasons why this prescription is generally not applicable beyond the leading order corrections. In subsection \ref{sec:interp} we give an interpretation of our formula in terms of a set of classical and quantum Planck-thickness bit threads. We continue in section \ref{sec:GeneralProps} where we discuss some general properties of our flow program, including nesting and a quantum version of the max multiflow theorem. These properties are then used to verify that our proposal respects known properties that a von Neumann entropy must satisfy, including subadditivity and strong subadditivity inequalities. Finally, we end the section by analyzing the fate of the monogamy of mutual information inequality when the leading $1/N$ corrections are included. Section \ref{sec:IyerWald} is devoted to the study of perturbative quantum states. Building up on our previous work \cite{Agon:2020mvu}, we show that the Iyer-Wald formalism can be used to provide a canonical flow configuration that solves the max-flow problem (even in its quantum version) which exploits the property of bulk locality. Combining with results by Swingle and Van Raamsdonk, we then show that this special solution requires the \emph{semi-classical} Einstein's equations to hold for any consistent perturbative bulk quantum state. Semi-classical gravity is then seen to arise consistently from entanglement considerations in the dual CFT. We close in section \ref{sec:Disc} with a brief summary of our results and a few final remarks.

\section{Flow program for quantum bit threads\label{sec:FlowProgram}}

When we consider quantum corrected versions of the RT formula, the standard bit thread construction should still hold true at leading order in $G_N$. For the sake of notation, then, we will use subscripts ``$(0)$'' in both sides of the equation (\ref{BTpres}) to indicate that these quantities do not include $G_N$ corrections:
\be\label{BTpres2}
S^{(0)}[A]=\frac{1}{4G_N}\, \max_{v^{(0)}\in \mathcal{F}}\int_{A} v^{(0)}\,,\qquad {\cal F}\equiv\{v^{(0)}\, \vert\, \nabla\cdot v^{(0)}=0,\, |v^{(0)}|\leq 1\}\,.
\ee
Notice that since $v^{(0)}$ is divergenceless we can chose to integrate over any other homologous region, in particular, over the bulk bottle-neck (or `min cut') $\gamma_A$ where $v^{(0)}$ must be equal to the unit normal $v^{(0)}=\hat{n}$. Therefore, one finds that
\be\label{BTpres2b}
S^{(0)}[A]=\frac{1}{4G_N}\int_{\gamma_A}\! dA=\underset{\gamma_A\sim A}{\text{min}}\left[\frac{A(\gamma_A)}{4G_N}\right]\,,
\ee
and one recovers the standard RT prescription. It is interesting to ask if the quantum corrections admit a similar description in terms of a corrected vector field $v=v^{(0)}+\delta v$, and if so, what are the corrections to the bit thread prescription.

In this section we will try to answer this question in a number of ways. We will begin with a heuristic analysis of the problem to determine how the corrections should look like. In this part we will assume the JLMS formula, an operator version of the FLM formula. Hence, our arguments will be valid for the leading quantum corrections. Next, we will provide a formal proof of our formula using convex optimization and strong duality. In this part we will assume the QES formula as a starting point, however, as we will explain, our derivation will only hold at order $\mathcal{O}(G_N^0)$. We will close the section by providing a physical interpretation of our prescription, as well as an analysis of quantum phase transitions.

\subsection{Heuristic derivation\label{sec:heuristic}}

\begin{figure}[t!]
  \centering
  \includegraphics[scale=0.3]{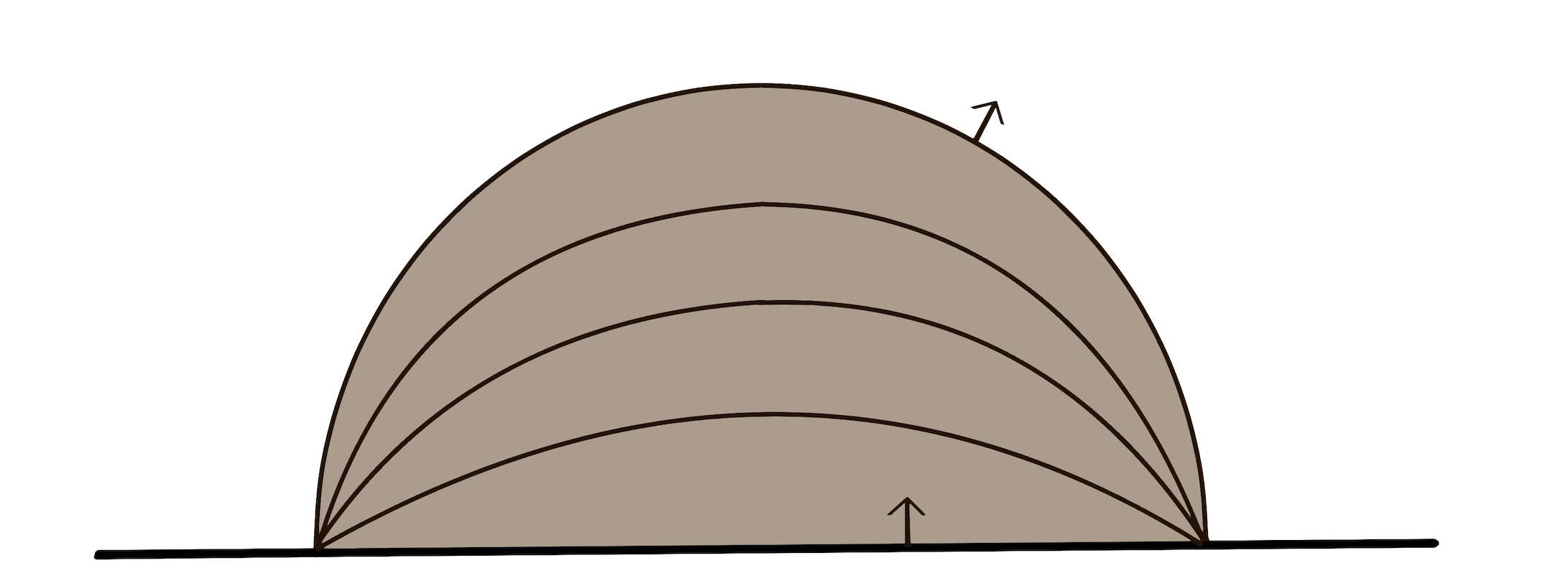}
  \begin{picture}(0,0)
\put(-196,15){$m_f=A$}
\put(-137,15){$\hat{n}_f$}
\put(-174,42){$\vdots$}
\put(-180,65){$m_3$}
\put(-180,87){$m_2$}
\put(-196,110){$m_1=\gamma_A$}
\put(-120,98){$\hat{n}_1$}
\put(-230,13){$\Sigma_A$}
\end{picture}
  \vspace{0mm}
  \caption{Family of bulk surfaces $m_i$ homologous to the boundary region $A$ that continuously interpolate between $\gamma_A$ and $A$, and hence foliate $\Sigma_A$.}\label{fig:contours}
\end{figure}
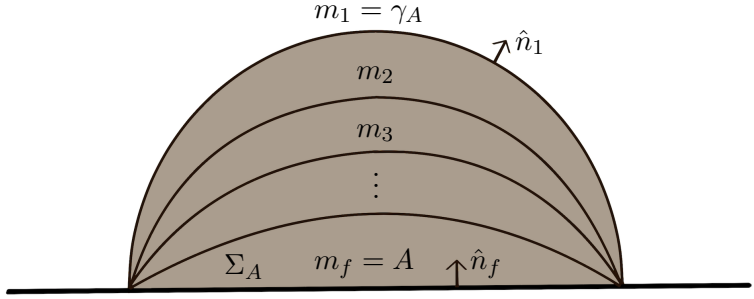

First, we note that a more refined version of the FLM formula states that \cite{Jafferis:2015del}
\be\label{JLMS}
\hat{K}_\text{CFT}=\frac{\hat{A}_{\text{min}}}{4G_N}+\hat{K}_\text{bulk}\,,
\ee
which now applies as an operator equation. In the above $\hat{A}_{\text{min}}$ is the minimal area operator, while $\hat{K}_\text{CFT}$ and $\hat{K}_\text{bulk}$ are the CFT and bulk modular Hamiltonians, respectively. We note that $\hat{K}_\text{CFT}$ has only support in $\partial \Sigma$ (specifically in $x \in D[A]$), while $\hat{K}_\text{bulk}$ has support in $\Sigma$ (concretely, in the entanglement wedge of $A$, $\{x,z\}\in D[\Sigma_A]$). In the following, we will take $\Sigma$ to be a constant-$t$ slice, for simplicity. Now, we can compute the flux of $v$ through a family of surfaces $m_i\sim A$ that continuously interpolate between $\gamma_A$ and $A$ (see Figure \ref{fig:contours}). As a result, we should obtain that the flux receives increasing contributions (either positive or negative) from the second term in (\ref{JLMS}). More concretely, if we compute the flux over $m_1=\gamma_A$, we should find that \emph{only} the area term contributes so that
\be\label{intgammaA1}
\frac{1}{4G_N}\int_{\gamma_A} v=\frac{1}{4G_N}\int_{\gamma_A} v^{(0)}=\frac{1}{4G_N}\int_{A} v^{(0)}=S^{(0)}[A]\,,
\ee
while if we compute the flux over the last surface $m_f=A$ we should expect, in addition, the \emph{full} contribution coming from the second term in (\ref{JLMS}),
\be\label{intA1}
\frac{1}{4G_N}\int_{A} v=\frac{1}{4G_N}\int_{A} (v^{(0)}+\delta v)=\frac{1}{4G_N}\int_{A} v^{(0)}+\frac{1}{4G_N}\int_{A} \delta v=S^{(0)}[A]+S_{\text{bulk}}[\Sigma_A]\,.
\ee
This is because at this point we have already swept over all the domain where $\hat{K}_\text{bulk}$ has support on (within $\Sigma$), i.e., $\Sigma_A$. In other words, we expect that
\be
\frac{1}{4G_N}\int_{A} \delta v=S_{\text{bulk}}[\Sigma_A]\,.
\ee
Recognizing that $\partial \Sigma_A=\gamma_A-A$,\footnote{The normal vectors are taken to be pointing towards the bulk (see Figure \ref{fig:contours} for an illustration).} we can use Gauss's law to write:
\be
\frac{1}{4G_N}\left(\int_{\gamma_A} v-\int_{A} v\right)=\frac{1}{4G_N} \int_{\Sigma_A} (\nabla\cdot v)=\frac{1}{4G_N} \int_{\Sigma_A} (\nabla\cdot \delta v)\,.
\ee
Using (\ref{intgammaA1})-(\ref{intA1}), we recognize that
\be
\frac{1}{4G_N} \int_{\Sigma_A} (\nabla\cdot \delta v)= -S_{\text{bulk}}[\Sigma_A]\,.
\ee
Thus, provided we can calculate an entanglement density in the bulk,\footnote{We can take $s(x)$ to be an entanglement \emph{contour} \cite{Chen_2014}. However, we do not need $s(x)$ be positive definite.} such that
\be
S_{\text{bulk}}(\Sigma_A)=\int_{\Sigma_A} s(x)\,,
\ee
it follows that
\be
\nabla\cdot \delta v = -4G_N\,s(x)\,.
\ee
For regions with a local modular Hamiltonian, we generally have that
\be
S_{\text{bulk}}=\langle \hat{K}_{\text{bulk}}\rangle=\int_{\Sigma_A}  \langle T_{\mu\nu}(x)\rangle  \xi^\mu {\bm \epsilon}^\nu\,,
\ee
where $T_{\mu\nu}$ is the \emph{bulk} stress-energy tensor, $\xi$ is a Killing vector with the right properties at $\gamma_A$, and ${\bm \epsilon}^\nu$ is the volume form on $\Sigma$. Hence, we can write
\be\label{density:localMH}
s(x)= \langle T_{\mu\nu}(x)\rangle \xi^\mu N^\nu\,,
\ee
and
\be
\nabla\cdot \delta v = -4G_N\,  \langle T_{\mu\nu}(x)\rangle \xi^\mu N^\nu \,,
\ee
where $N^\mu$ is the future-pointing unit normal associated with $\Sigma$. Notice that the modular Hamiltonian has only support on $\Sigma_A$. However, we can extend this condition to all the slice $\Sigma$ by analytically continuing the bulk modular Hamiltonian to the complement $\bar{\Sigma}_A=\Sigma\backslash\Sigma_A$.

A couple of comments are in order. First, since the divergenceless condition is violated, it means that now we can have ``sources'' and ``sinks'' in the bulk entering at order $\mathcal{O}(G_N)$. Both are \emph{necessary} in our prescription. For example, for a pure state we expect $S_{\text{bulk}}=0$ and hence the sources and sinks should be in exact balance.\footnote{One can always set $s(x)=0$ for pure states, but this will severely restrict the microstate. In more general cases, this will not be an option. See section \ref{sec:interp} for a more thorough discussion on this point.} The relaxation of this condition is, nevertheless, nice for the interpretation, because now the threads can also start and end in the bulk, and we could now distill bell pairs from the bulk as well. On the other hand, this implies that if we assume effective field theory in the bulk, a bulk observer can \emph{only} extract bell pairs from the state, and no other form of multipartite entanglement (c.f. section 6.5 of \cite{Jafferis:2015del}). Finally, the norm bound can in principle be violated since there is no general bound for $s(x)$,  although we expect $s(x)=0$ at the bulk bottle-neck, $\gamma_A$, where $|v^{(0)}|$ is saturated. The possible violation would appear close to $\gamma_A$, where $|v^{(0)}|$ is still close to saturation, but it would be of order $\mathcal{O}(G_N)$ at most.
Happily, as we will show in the next section, it turns out there is no need for corrections to the norm-bound. This can be explicitly checked as our proposal can be formally derived using convex optimization techniques.

\subsection{Proof via convex optimization\label{ConvexOpt}}

Having discussed the expected leading correction to the bit thread prescription of holographic entanglement entropy, we will now formally derive the quantum corrected flow program for the applicable set of CFT states with semi-classical bulk geometries. For the derivation of the dual program we will assume that, given a bulk microstate or a family of microstates, we have a way of obtaining an entanglement density $s(x)$, such that
\bea\label{Sbulk}
S_{\text{bulk}}[\Sigma_A]=\int_{\Sigma_A}s(x)\,.
\eea
A few comments are in order. First notice that  to construct one of these densities, one needs as input knowledge of the homology region $\Sigma_A$ associated to the boundary region $A$, and its associated bulk entanglement entropy $S_{\rm bulk}[\Sigma_A]$. These are already the ingredients one needs to compute the quantum corrections to entanglement entropy. However, the main purpose of the flow reformulation we will arrive at is to provide conceptual understanding of the FLM formula, rather than to provide an independent calculational tool of entanglement entropies. Second, if the reader is worried about uniqueness of the function $s(x)$, note that, for simplicity and in order to completely specify the program, one \emph{could} assume that $s(x)\geq 0$ for $x\in\Sigma_A$ so that $s(x)$ defines a proper entanglement contour \cite{Chen_2014,Wen:2018whg} in the bulk, for the reduced state on $\Sigma_A$. This is, however, not a requirement for our prescription to work. In fact, in many cases, it will be necessary to relax this assumption and let $s(x)$ take both positive and negative values. We will come back to this point in the next section, where we prove several properties of quantum bit threads.

The starting point of the proof is the equivalence between an \emph{area} minimization problem and a \emph{volume} minimization problem. In \cite{Headrick:2017ucz} it was argued that, at the classical level, the minimization in the RT prescription is equivalent to:
\bea
\min_{\gamma_A\sim A}\left[ \frac{{A}(\gamma_A)}{4 G_N}\right] =\min_{\phi,\phi|_{\partial \Sigma}=\chi_{_A}} \left[\frac{1}{4G_N}\int_{\Sigma}\! |\partial \phi|\right]\,.
\eea
The latter minimization is carried out over bulk scalar fields $\phi$ satisfying the boundary condition $\phi|_{\partial \Sigma}=\chi_{_A}$, where $\chi_{_A}=1$ on $A$ and $0$ otherwise. Further, \cite{Headrick:2017ucz} showed that the minimum of the RHS is generically achieved by a step like function $\phi_{\text{min}}=\chi_{_{\Sigma_A}}$, with $\chi_{_{\Sigma_A}}=1$ on $\Sigma_A$ and $0$ otherwise. Following the same logic, we can now argue that the QES prescription is equivalent to the following volume minimization:
\be\label{eq:volumeQES}
S[A]=\min_{\phi, \phi|_{\partial M}=\chi_{_A}} \left[\frac{1}{4G_N}\int_{\Sigma}\! |\partial \phi| +\int_{\Sigma}\! \phi\,s(x) \right]\,.
\ee
where we have made use of the local expression for the bulk entropy (\ref{Sbulk}). One crucial point here is that we have assumed that the density $s(x)$ is given as an input of the program. This cannot be done in general, as there is no density that can compute the entanglement entropy of arbitrary regions. The best we can do here is to assume that the bulk entropy is parametrically smaller than the area term, and hence we have countable possible saddles for $\Sigma_A$. These possible saddles will be nested, in general, so even if we have more than one classical solution, it will still be possible to come up with a sensible density $s(x)$ defined everywhere. Another reason why we cannot deal with the full expression (\ref{eq:volumeQES}) is that, we cannot generally let the minimization backreact on the density $s(x)$. If we allow for this possibility, it will render the final program non-convex (more on this below) which means that the techniques that we use would break down at some point.\footnote{Double holographic setups are an exception to this rule. In these cases, a version of our results should hold, because the dual program can be mapped to that of a divergenceless flow in a higher dimensional space. We will comment on this possibility in the discussion section.} In the situations that we have under control, then, it is valid to expand (\ref{eq:volumeQES}) at leading order to obtain
\bea
S[A]&\approx&\min_{\phi, \phi|_{\partial M}=\chi_{_A}} \left[\frac{1}{4G_N}\int_{\Sigma}\! |\partial \phi|\right]  +\int_{\Sigma}\! \chi_{_{\Sigma_A}} s(x) \,,\\
&\approx& \min_{\phi, \phi|_{\partial M}=\chi_{_A}} \left[\frac{1}{4G_N}\int_{\Sigma}\! |\partial \phi|\right]  +\int_{\Sigma_A}\!s(x)\,.\label{mincut-FLM}
\eea
This volume minimization is equivalent to the FLM formula.  As mention in the introduction, however, there is a small subtlety because the QES and FLM formulas can differ at order $O(G_N^0)$ in situations close to a phase transition. For this reason, will continue working with (\ref{eq:volumeQES}), though, keeping in mind that our results will only be sensible at order $O(G_N^0)$.
Another advantage of (\ref{eq:volumeQES}) is that, written in this way, we can now massage the expression into a max flow program using the same techniques developed in \cite{Headrick:2017ucz}, which will be our main goal for the rest of this section.

To start the proof, notice that (\ref{eq:volumeQES}) defines a convex program. The variable is the scalar field $\phi$ and the objective in (\ref{eq:volumeQES}) is a convex functional of $\phi$. There is an equality constraint which is affine on $\phi$ (linear plus constant), namely $\phi|_{\partial \Sigma}=\chi_A$ and then it is trivially convex as well. In order to obtain the dual max flow program we now introduce a co-vector like variable, $w_\mu=\partial_\mu\phi$, and Lagrange multipliers, $v^\mu$ and $\psi$, which impose the implicit constraints in the following way (factorizing an overall factor of $1/4G_N$):
\bea
\int_{\Sigma}\! |\partial \phi| +4G_N\!\int_{\Sigma}\! \phi\,s(x)=\max_{v^\mu, \psi}  \left[\int_{\Sigma}\! |w|+4G_N\int_{\Sigma}\! \phi\, s(x) +\int_{\Sigma}\!v^\mu\(w_\mu-\partial_\mu \phi\)+\int_{\partial \Sigma}\!\!\psi \(\phi-\chi_A\)\right]\!.\nonumber
\eea
The way in which the constraints emerge from the above maximization is by imposing the condition of finiteness. This forces $w_\mu \to \partial_\mu \phi$ and $\phi|_{\partial \Sigma}\to \chi_A$. The dual program is then obtained by inverting the order of maximization and minimization steps, i.e.,
 \bea
 \!\!\!\!\!\!\!\!\!S[A]=\frac{1}{4G_N}\max_{v^\mu, \psi}\left\{\min_{\phi, w_\mu}\left[\int_{\Sigma}\! |w|+4G_N\!\int_{\Sigma}\! \phi\, s(x) +\int_{\Sigma}\!v^\mu\(w_\mu-\partial_\mu \phi\)+\int_{\partial \Sigma}\!\!\psi \(\phi-\chi_A\)\right]\right\}\!.
 \eea
Since we turned the implicit constraint on $\phi$ into an explicit one via the Lagrange multipliers, $\phi$ and $w_\mu=\partial_\mu \phi$ should be treated as independent variables. As a result, the minimization is now over both $\phi$ and $w_\mu$.
Rearranging the terms and integrating by parts, we obtain:
\bea
 \!\!\!\!\!\!\!&&\min_{\phi, w_\mu}\left[\int_{\Sigma}\!\( |w| +v^\mu w_\mu \)- \int_{\Sigma}\!\nabla_\mu \(v^\mu \phi\) +\int_{\Sigma}\! \phi \nabla_\mu v^\mu+\int_{\partial \Sigma}\!\!\psi \(\phi-\chi_A\)+4G_N\!\int_{\Sigma}\! \phi\,s(x) \right] \nonumber \\
 \!\!\!\!\!\!\!&=& \min_{\phi, w_\mu}\left[\int_{\Sigma}\!\( |w| +v^\mu w_\mu \)+\int_{\partial \Sigma}\!\! \phi\(v^\mu n_\mu +\psi\)-\int_{\partial \Sigma}\!\! \psi \chi_A + \int_{\Sigma}\! \phi\, \( \nabla_\mu v^\mu+4G_Ns(x)\)\right]  \nonumber \\
 \!\!\!\!\!\!\!&\geq & \min_{\phi, w_\mu}\left[\int_{\Sigma}\!|w| \( 1-|v|\)+\int_{\partial \Sigma}\!\! \phi\(v^\mu n_\mu +\psi\)-\int_{\partial \Sigma}\!\! \psi \chi_A + \int_{\Sigma}\! \phi\, \( \nabla_\mu v^\mu+4G_Ns(x)\)\right].
 \eea
The first equality is obtained by using the divergence theorem to integrate a total derivative. The inequality is derived from a bound on the first term which is achieved by picking an optimal direction for $w_\mu$, such that $v^\mu w_\mu=-|v|\,|w|$. This is achieved by minimizing over the direction of $w_\mu$, and thus sets a lower bound indicated by the inequality.

Finiteness of the above minimization problem requires the following constraints on the dual variables $\psi$ and $v^\mu$. The first term implies the norm bound
\be\label{eq:normb}
|v|\leq 1\,,
\ee
in which case such term contributes zero to the dual objective functional. Similarly, the second terms requires $\psi=-v^\mu n_\mu$ and likewise contributes zero to the dual objective functional. The fourth term implies a new constraint on $v$
\bea\label{eq:constr}
\nabla\cdot v=-4G_N s(x)\,,
\eea
which when evaluated in the objective functional gives a zero contribution as well. Putting everything together we see that the only contribution to the dual objective functional comes from the third term, which reduces to:
\bea
\int_A v^\mu n_\mu\,,
\eea
and therefore the quantum corrected dual program is given by:
\bea\label{eq:flowFLM}
S[A]&=&\max_{v \in \mathcal{F}}\int_A v\,,\qquad \mathcal{F}\equiv \{v\, \vert\, \nabla\cdot v=-4G_N s(x),\,|v|\leq 1 \}\,,
\eea
where $s(x)$ obeys (\ref{Sbulk}). This concludes the proof of our quantum bit thread prescription.

Under sensible assumptions, we expect the solution of this dual program to be equal to that of the primal, as a result of \emph{strong duality}. One simple condition that implies strong duality is Slater's condition \cite{BookConvex}, which states that there should be a flow (not necessarily the maximal one) which strictly satisfies all the equality and inequality constraints. This is an easy task to accomplish, at least at the order of approximation that we are working on. To see this, notice that we can split a given flow $v$ in two pieces, a homogeneous and an inhomogeneous part, such that
\be\label{eq:flowsHI}
v=v_{\text{h}}+v_{\text{i}}\qquad\text{with}\qquad \nabla\cdot v_{\text{h}}=0\,,\qquad \nabla\cdot v_{\text{i}}=-4G_N s(x)\,.
\ee
Thus, the equality constraint (\ref{eq:constr}) only affects the inhomogeneous piece $v_{\text{i}}$. However, since the source term enters at order $\mathcal{O}(G_N)$, there must be particular solutions that are also of the same order and so their flux through any macroscopic region $A$ will not pose any obstacle for the inequality constraint, or norm bound (\ref{eq:normb}), which only enters at order $\mathcal{O}(G_N^0)$.\footnote{Consistency of the problem requires restricting to densities $s(x)$ which are at most order $\mathcal{O}(G_N^0)$ in any local neighborhood. If the density is higher than this, say $s(x)\sim\mathcal{O}(1/G_N)$, we could risk violations of the norm bound already with $v_i$. This restriction does not pose any threat to the program as it can always be satisfied in situations relevant to us, i.e., when the bulk entropy is parametrically smaller than the area term.\label{footnote-maxdensity}} We can then pick any flow with $v_{\text{h}}\sim \mathcal{O}(G_N)$ and $v_i\sim \mathcal{O}(G_N)$ satisfying (\ref{eq:flowsHI}) and it will be an example of a flow that lies is in the interior of the program's domain, hence, implying strong duality. We note that, once the equality constraint is satisfied, a solution of the max flow program can then be found by increasing the flux of the homogeneous solution, possibly to order $\mathcal{O}(G_N^0)$, until the norm bound is saturated.

As a final comment, we remind the reader that our quantum bit thread program is only valid when the leading order corrections are taken into account, i.e., at order $\mathcal{O}(G_N^0)$,\footnote{One way to see that this prescription is invalid beyond leading order is by noticing that the equality constraint (\ref{eq:constr}) becomes non-convex if we let the flux maximization to backreact on $s(x)$. This should not be possible because we assumed the density $s(x)$ to be given in order to define the program. Another obstacle that arises beyond leading order is the impossibility of generally satisfying Slater's condition, discussed above.} though it goes a step further than the FLM prescription. In particular, unlike FLM, our formula is valid arbitrarily close to phase transitions, at $\cO(G_N^0)$, picking the dominant configuration among all possible classical saddles.\footnote{The FLM formula can fail at choosing the right saddle as discussed in detail in section \ref{QPT}.} In the next section we will explore some of the consequences of this reformulation and prove a set of properties that this proposal must satisfy.

\subsection{Physical interpretation\label{sec:interp}}

The flow program that we derived (\ref{eq:flowFLM}) is corrected with respect to the standard one (\ref{BTpres}) in that we can now have ``sources'' and ``sinks'' in the bulk entering at order $\mathcal{O}(G_N)$. This means that threads will now have the possibility of start and/or end in the bulk. We will refer to those threads as \emph{quantum bit threads}. Quantum bit threads codify a new class of distillable entanglement, now present in the bulk itself. The interplay between classical and quantum bit threads is subtle, though, as sometimes they could be interpreted in one way or another depending on the bulk regions that we consider or have access to. To illustrate this point, we can split a given solution to the program in two pieces, a homogeneous and an inhomogeneous part, as in (\ref{eq:flowsHI}).
Upon discretization, then, one can be tempted to interpret the threads that follow from $v_{\text{i}}$ as quantum threads. However, we will see that in some cases the flux computed from $v_{\text{i}}$ actually contributes to the area piece and not to the bulk entanglement. This is due to the fact that such a splitting (\ref{eq:flowsHI}) is not unique, and one can always add part of the homogeneous solution to the inhomogeneous one. The goal of this section is to dig further into this observation. We will consider two illustrative cases, with one or multiple (classical) saddle points, respectively. We will close the section analyzing the role of quantum bit threads in possible phase transitions between different saddles. This analysis also highlights the distinction between the FLM and QES formulas at order $\mathcal{O}(G_N^0)$, as well as the correct interpretation of the dual program.

\subsubsection{Possible classical saddles}

\subsubsection*{Unique saddle:}
\begin{figure}[t!]
\centering
 \includegraphics[width=2.7 in]{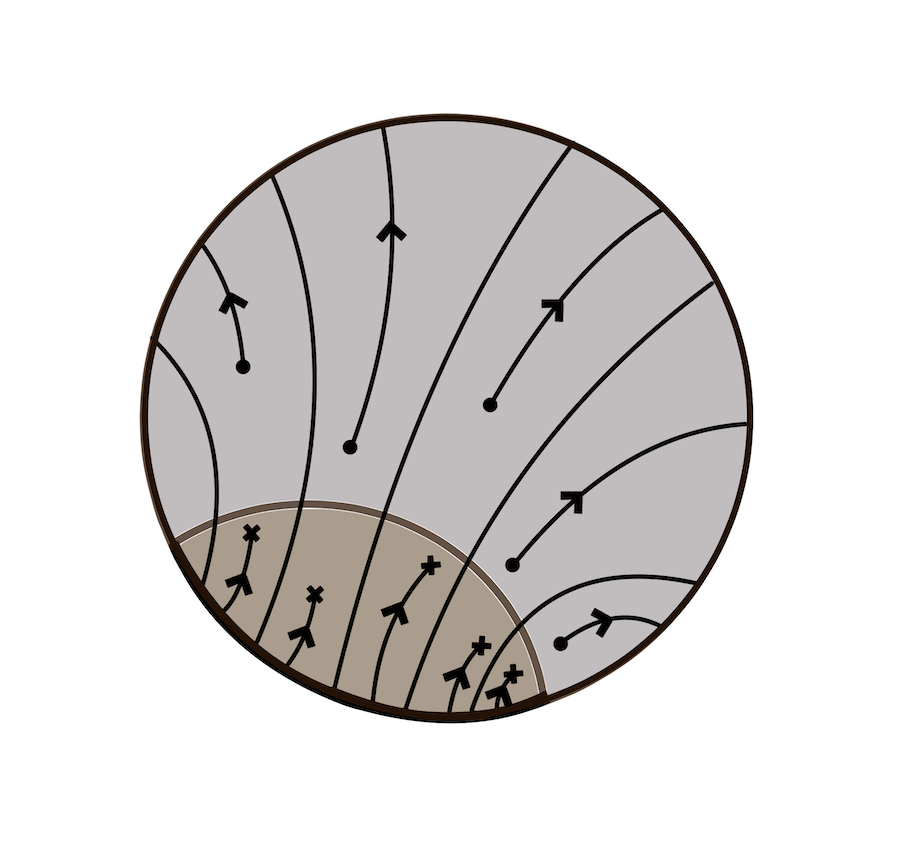}$\qquad$ \includegraphics[width=2.7 in,trim=0 -0.7in 0 0]{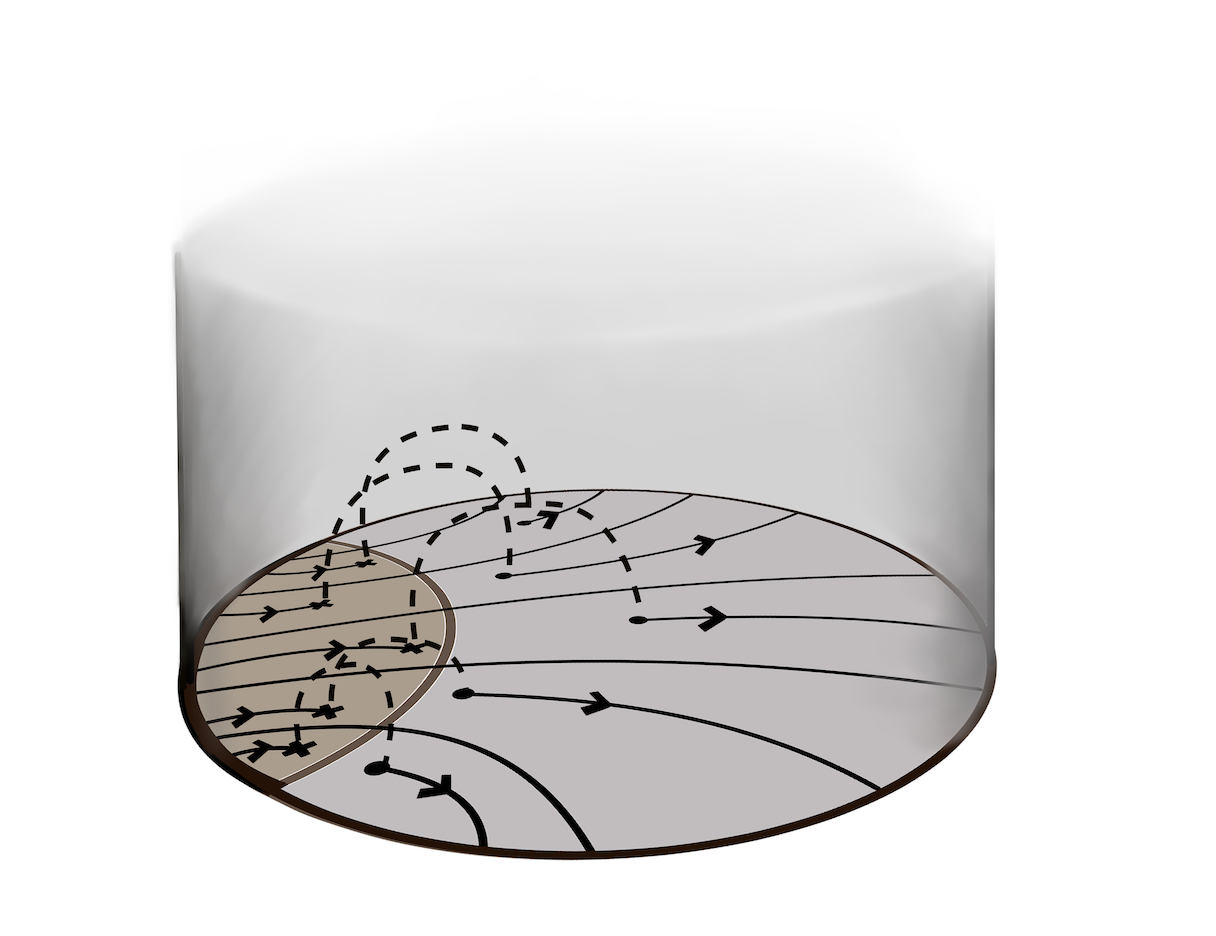}
   \begin{picture}(0,0)
\put(-360,21){$A$}
\put(-351,61){$a$}
\put(-304,157){$\bar{A}$}
\put(-323,78){$\bar{a}$}
\put(-181,55){$A$}
\put(-37,55){$\bar{A}$}
\put(-36,85){$\uparrow$}
\put(-36.4,95){$u$}
\end{picture}
 \caption{Thread representation of the max flow associated to a simply connected boundary region (left). The crosses represent sinks of threads while the dots represent sources. Quantum threads are thus naturally oriented objects. From the global perspective, they are seen to \emph{jump} across the minimal surface via quantum tunneling. This process can be intuitively understood in double holographic models (right), where the quantum threads are still continuously connected through the an emergent coordinate $u$. In this context, the sinks and sources arise by a projection of a divergenceless field into a lower dimensional space. \label{fig1}}
 \end{figure}
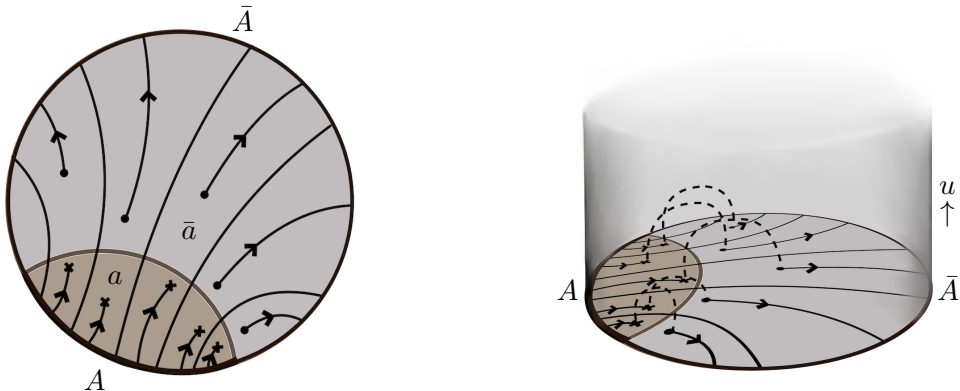

Let us first consider the case where the minimal surface $\gamma_A$ is unique and so is the bulk homology region associated to it $\Sigma_A$. In this case we can
pick a density function $s(x)$ with a definite sign on $\Sigma_A$ and $\bar{\Sigma}_{A}$ (opposite to each other), which is possible given \emph{any} choice of the bulk state \cite{Chen_2014}. From the point of view of the bulk subregions, $\Sigma_A$ and $\bar{\Sigma}_{A}$, then, there will be a natural separation between threads that compute the minimal area and the ones that compute the bulk entropies. More specifically, the threads that connect $A$ with $\bar{A}$ compute the minimal area contribution while the threads that start or end in the interior of the bulk compute the entropies associated with $\Sigma_A$ and $\bar{\Sigma}_{A}$, respectively. These two bulk entropies must be the same if the overall state is pure. This implies that the  total number of threads leaving $A$, both classical and quantum, must be the same number of threads entering $\bar{A}$. From the global perspective, then, the quantum threads can be interpreted as those that \emph{jump} across the minimal surface, e.g., via a tunneling process, but still connect the boundary regions $A$ and $\bar{A}$. In double holographic scenarios such tunneling can be realized geometrically by assuming that those threads are continuously connected in the higher dimensional holographic dual (effectively realizing a holographic `EPR pair' in the higher dimensional space. See e.g. \cite{Jensen:2013ora,Sonner:2013mba,Chernicoff:2013iga}). Figure \ref{fig1} gives a pictorial representation of one of these thread configurations. We emphasize that, in this case there is always the option of choosing a density with the above properties, though, more generally, we \emph{can} allow $s(x)$ to change signs within $\Sigma_A$. In this case the interpretation would be a bit more complicated. The reason is that if we allow for this possibility there would be threads that connect points in $\Sigma_A$ with $\bar{A}$ (hence they would naively fall in the category of quantum threads) yet they contribute exclusively to the area term. This means that already in cases with a unique saddle, there can be cases where quantum threads are not interpreted in the standard way. If there are multiple saddles, on the other hand, this situation could be unavoidable. As we will see below, in these cases we do not have always the option of picking a density with the desired properties, which means that these threads will naturally arise in more general situations.

\subsubsection*{Multiple saddles:\label{multiple-ms}}
\begin{figure}[t!]
\centering
 \includegraphics[width=3.4 in]{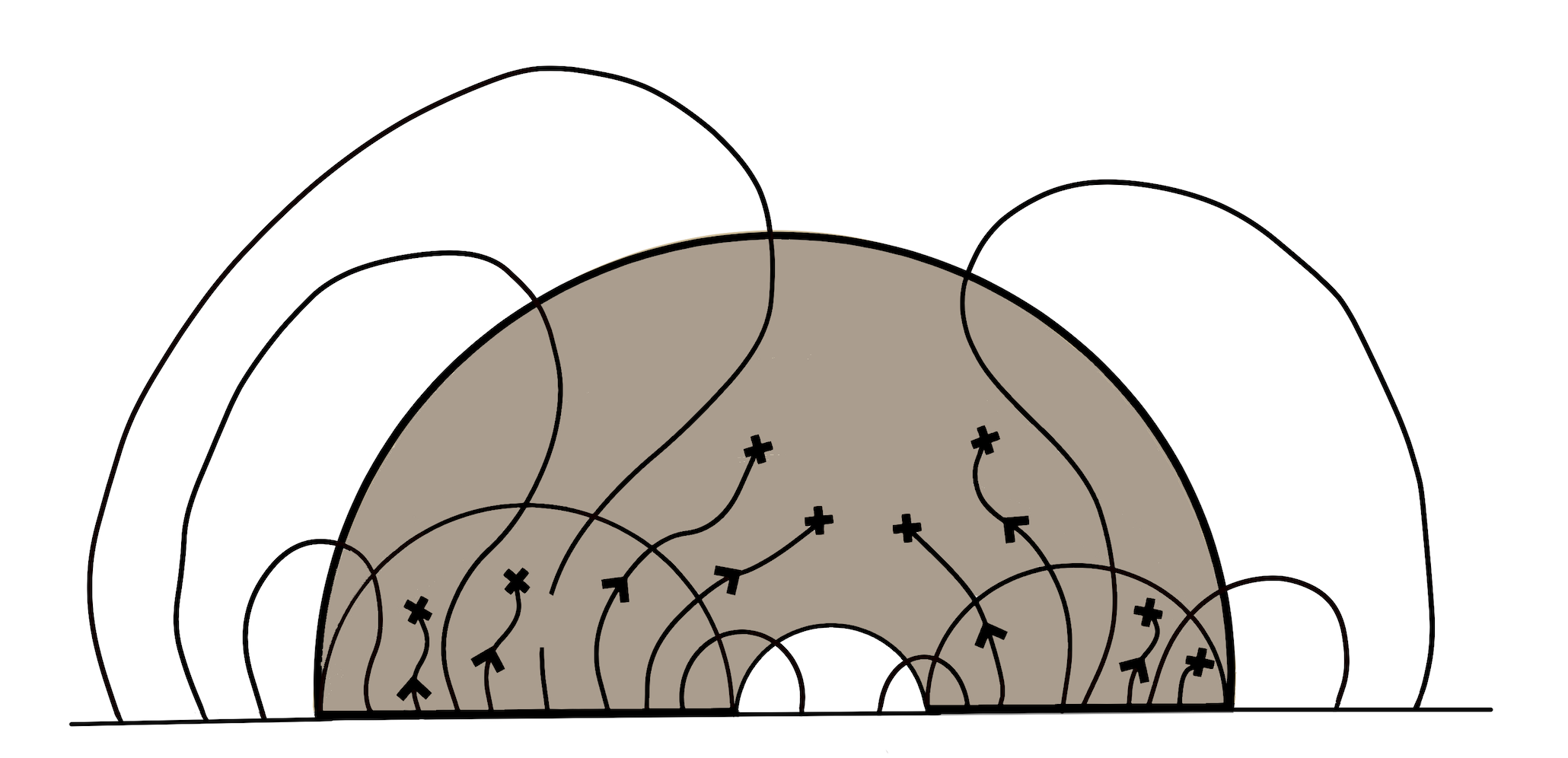}
   \begin{picture}(0,0)
\put(-170,-5){$A$}
\put(-85,-5){$B$}
\put(-167,19){$a$}
\put(-115,45){$c$}
\put(-90,15){$b$}
\end{picture}
  \vspace{0mm}
 \caption{Thread representation of the max flow associated to a boundary region made out the two simply connected parts $A$ and $B$. Above, we have defined $a$, $b$ and $ab$ as the bulk homology regions associated with $A$, $B$ and $AB$, respectively, and $c=ab\backslash a\cup b$. The entropy density on $a$, $b$ and $c$ is picked to be positive and thus acts as sinks for the quantum threads. When restricted to the region $a\cup b$, though, some of these threads (the ones ending on $c$) are interpreted as classical because they contribute exclusively to the area $\partial \(a\cup b\)$. 
 \label{fig2}}
 \end{figure}
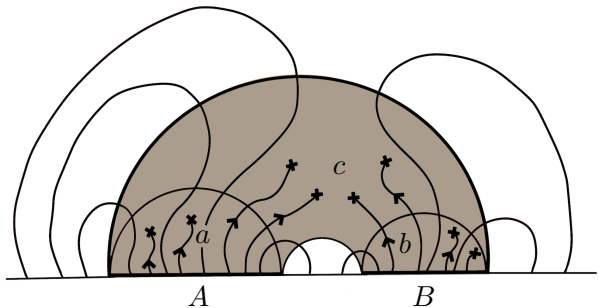

It is well known that the minimal surface associated to a boundary region can undergo a phase transition (change in topology) under continuous deformations of the boundary geometry. The prototypical example consists of two disjoint regions with variable separation ---see Figure \ref{fig2} for an illustration---. Perhaps less known is that, already at order $\mathcal{O}(G_N^0)$, the bulk entropy \emph{can} induce transitions on the corresponding minimal surfaces if the configuration of the boundary region is sufficiently close to a classical phase transition. This is a crucial distinction between the QES and FLM prescriptions which already exists at this order. As we will see below, our quantum bit thread prescription knows about this subtlety, and correctly captures the result expected from the QES formula.

As a first step, let us consider the exact point of the classical transition. In this case there are multiple minimal surfaces $\gamma_A$'s, all with the same area but with different associated homology regions $\Sigma_A$'s. In order to use our quantum bit thread prescription, it is convenient to pick a bulk density that is able to reproduce the entropy of all the possible bulk homology regions. For concreteness, let us assume that there are $N$ possible homology regions
 and let us label the $\Sigma_A$'s as $\{a^i\}$ with $i\in \{1, \cdots, N\} $. Consistency with subregion duality implies that these bulk regions must be nested\footnote{The heart of the argument is the following. Suppose that, at a transition point, there are two equally dominant homology regions $a_1$ and $a_2$ associated with a boundary region $A$, which are not nested. Next, consider an infinitesimal deformation of the region, $\tilde{A}=A+\delta A$, such that $\tilde{A}\supset A$. This new region is away from the exact transition point, making one of the homology regions dominate, say $\tilde{a}_1$. If that is the case, then, nesting between regions $A$ and $\tilde{A}$ would imply that $\tilde{a}_1\supset a_1$ and $\tilde{a}_1\supset a_2$. However, in the limit $\delta A\to0$, by continuity, we know that one of these two statements must be wrong, reaching a contradiction. This implies that there should be another classical saddle, $a_3$, that dominates over $a_1$ and $a_2$ at the transition point. Alternatively, this argument implies that all relevant saddles at a transition point must be nested.}, and thus we can label them from smaller to larger such that $a^i\subset a^{i+1}$ for $i\in \{1, \cdots, N-1\} $. Alternatively, since these regions are different by assumption, one can define $N$ non-overlapping regions $a^1$ and $a^{i}\backslash a^{i-1}$ for $i\in \{2, \cdots, N\} $, and thus write the $k$th region, $a^k$, as the union of all previous regions,
\bea
a^k=a^1\cup(a^2\backslash a^1)\cup \cdots \cup (a^{k}\backslash a^{k-1})\,.
\eea
Now, we construct an entanglement density $s(x)$  such that it reproduces the bulk entropies $S[a^i]$ when integrated over each region $a^i$,
\bea
S[a^i]=\int_{a^i}s,\, \quad{\rm for\,\,  all \,\, } i\,.
\eea
We can achieve this by imposing the the following constraints:
\bea
\int_{a^1}s=S[a^1]\quad{\rm and}\quad \int_{a^{i+1}\backslash a^i}s=S[a^{i+1}]-S[a^{i}]\quad {\rm for \,\, all \,\, }i\in\{1, \cdots , N-1\}\,.
\eea
In other words, we assign the value of an entropy difference to the integrated density of each non-overlapping region, which can always be done. In this case, the quantum bit thread program (\ref{eq:flowFLM}) picks among all the possible homology regions the one that is favored by the maximality condition. Notice that this is in stark contrast to the FLM prescription, where all the possible saddles are equally valid even though their bulk entropies may differ. This implies that the quantum bit threads formalism captures correctly the answer that follows from the QES formula at the leading order in $G_N$. In terms of interpretation, we note that the label as quantum or classical bit threads
can only be specified with reference to a specific homology region. To give an example, consider threads that start in the region $A$ and end at some point within $a^{k}$ for $k>1$. Clearly, from the point of view of this bulk region, these threads are interpreted as quantum threads, given the definition introduced above. However, if the program picks $a^j$ with $j<k$ as the relevant homology region, then these threads are interpreted as classical. An example of this subtle ambiguity is presented in Figure \ref{fig2}.

\subsubsection{Quantum phase transitions \label{QPT}}

Let us now investigate in more detail the fate of the phase transition when the effect of the bulk entropy is taken into acocunt. For concreteness, we will consider the simplest example, a boundary region $AB$ made out of two disjoint intervals $A$ and $B$. We will assume that the configuration is \emph{close} to the classical phase transition, though, not exactly at the critical point. We will now illustrate thread configurations that compute the holographic entanglement entropy in various possible scenarios.

As it is well know, for a subsystems made out of two disjoint intervals there are two possible bulk homology regions.  These regions are labeled as $ab$ and $a\cup b$,\footnote{Or, in terms of the previous notation, $(ab)^{1}=a\cup b$ and $(ab)^2=ab$.} with $ab\supset a\cup b$, and correspond to the connected and disconnected configurations, respectively. The dominant saddle here will depend on the specific value of the bulk entropies and the respective minimal area surfaces, and the specific point of the transition would be such that
\bea\label{PT-point}
\frac{{\rm area}\( \partial ab\)}{4G_N}+S[ab]=\frac{{\rm area}\( \partial \(a\cup b\)\)}{4G_N}+S[a\cup b]\,.
\eea
As before, we define the density function $s(x)$ with the following properties in mind:
\bea\label{PT-s}
\int_{a\cup b}s=S[a\cup b]\quad {\rm and }\quad \int_{ab\backslash a\cup b}s=S[ab]-S[a\cup b]\,.
\eea
The bulk entropies $S[ab]$ and $S[a\cup b]$ are positive definite and thus one can generically choose positive densities for these regions. However, in order to use the quantum bit thread prescription we must provide a single density $s(x)$ from which we can reproduce both entropies. For cases with $S[ab]<S[a\cup b]$, this forces $s(x)$ to be negative in some regions as is implied by the second equation in (\ref{PT-s}). This gives rise to two possibilities which we will now study separately.

\subsubsection*{Case I: $S[ab]>S[a\cup b]$}

In this case, the transition equation (\ref{PT-point}) implies that the difference in the extremal area surfaces exactly compensate for the difference in bulk entropies
\bea\label{difareasI}
\frac{{\rm area}\( \partial \(a\cup b\)\)}{4G_N}-\frac{{\rm area}\( \partial ab\)}{4G_N}=S[ab]-S[a\cup b]>0\,.
\eea
In order for this to happen, the difference in areas should be of order $\mathcal{O}(G_N)$. This difference in areas implies that the configuration with homology region $ab$ would dominate classically, however, at order $\mathcal{O}(G_N^0)$ they both are equally relevant. Exactly at this point, there should be thread configurations that can compute the entropy of both homology regions. One of such configurations is depicted in Figure \ref{fig2}. We note that the flux through $AB$ associated with the collections of threads that end on $c=ab\backslash a\cup b$ equals $S[ab]-S[a\cup b]$. From the point of view of the $ab$ system, this flux computes bulk entropy. However, from the point of view of the $a\cup b$ system, this flux contributes to the area term. Indeed, from (\ref{difareasI}) we can deduce that this flux must give the excess in area, i.e., ${\rm area}\(\partial \(a\cup b\)\)/4G_N-{\rm area} \(\partial ab\)/4G_N$. Then, we conclude that by looking at either homology region, $ab$ or $a\cup b$, one can interpret different thread bundles as classical or quantum. From the global perspective, quantum threads compute bulk entropy, while if we focus on a bulk (nested) subregion, the same threads can compute area and, hence, can be interpreted as classical.

We can now move away from the exact phase transition point by slightly changing the state and hence the bulk entropies. This would make one of the two configurations to dominate and thus only one of the two sides of (\ref{PT-point}) would be computed via a thread configuration such as the one shown in Figure \ref{fig2}. We will study both possibilities next.

\begin{itemize}
  \item \emph{Disconnected solution dominates:} The disconnected solution would dominate if
\bea\label{dc-dom}
\frac{{\rm area}\( \partial \(a\cup b\)\)}{4G_N}-\frac{{\rm area}\( \partial ab\)}{4G_N}<S[ab]-S[a\cup b]\,.
\eea
This includes the case when the disconnected solution dominates classically, ${\rm area}\( \partial \(a\cup b\)\)<{\rm area}\( \partial ab\)$, as the above inequality would be trivially satisfied.\footnote{The LHS would be negative while the RHS is positive by assumption.} More interestingly, this also includes the case when the \emph{connected} solution dominates classically, ${\rm area}\( \partial ab\)<{\rm area}\( \partial \(a\cup b\)\)$, provided that the difference in areas obeys the bound (\ref{dc-dom}).
  \item \emph{Connected solution dominates:} The connected solution would dominate if
\bea\label{cc-dom}
\frac{{\rm area}\( \partial \(a\cup b\)\)}{4G_N}-\frac{{\rm area}\( \partial ab\)}{4G_N}>S[ab]-S[a\cup b]\,.
\eea
This only happens when the connected solution dominates classically,
 ${\rm area}\( \partial \(a\cup b\)\)>{\rm area}\( \partial ab\)$, by an amount determined by the entropy difference in (\ref{cc-dom}).
\end{itemize}

\subsubsection*{Case II: $S[ab]<S[a\cup b]$}

\begin{figure}
\centering
 \includegraphics[width=3.8 in]{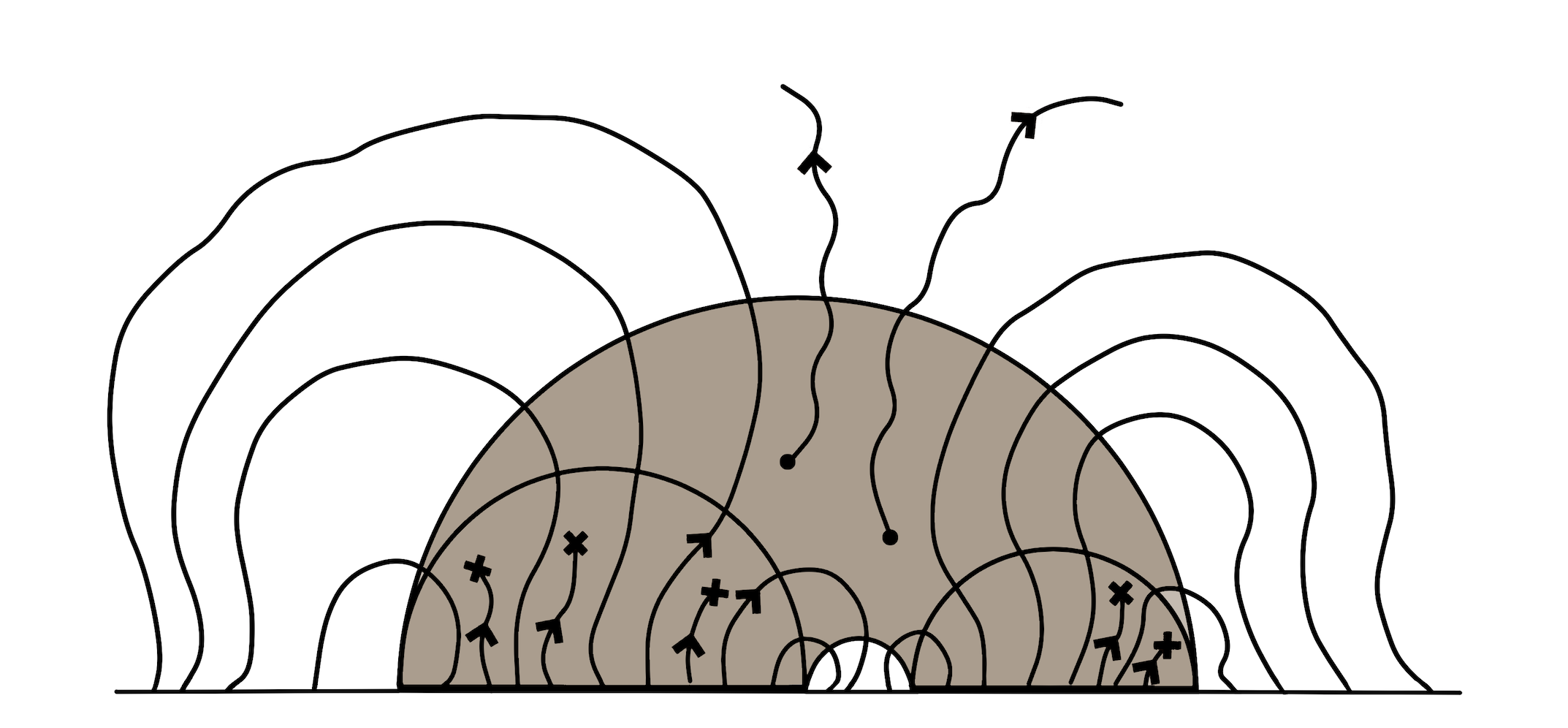}
    \begin{picture}(0,0)
\put(-177,-5){$A$}
\put(-97,-5){$B$}
\put(-171,19){$a$}
\put(-138,35){$c$}
\put(-104,15){$b$}
\end{picture}
  \vspace{0mm}
 \caption{Thread representation of the max flow associated to a boundary region made out of two simply connected parts $A$ and $B$. In this case, the entropy density on $c=ab\backslash a\cup b$ is negative and thus acts as sources for the quantum threads. When restricted to the region $ab$, though, some of these threads (the ones starting on $c$) are interpreted as classical because they contribute exclusively to the area $\partial \(a\cup b\)$. \label{fig3}}
 \end{figure}
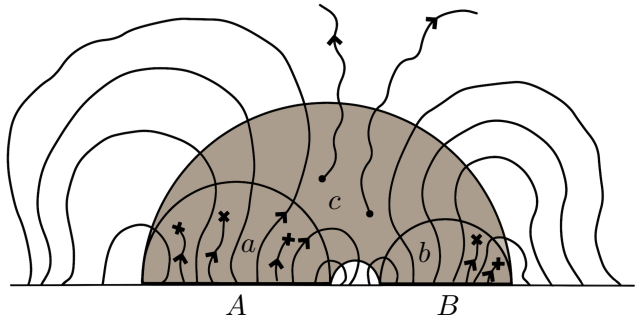

In this case, the transition equation (\ref{PT-point}) implies that
\bea
\frac{{\rm area}\( \partial ab\)}{4G_N}-\frac{{\rm area}\( \partial \(a\cup b\)\)}{4G_N}=S[a\cup b]-S[ab]>0\,.
\eea
This means that, at the classical level, the configuration with homology region $a\cup b$ would be the relevant one. However, at the quantum level, both are equally dominant. At the exact transition point there should be thread configurations that can compute the entropy of both saddles. In Figure \ref{fig3} we represent one such configurations. We note that the flux through $AB$ associated with the collections of threads that start on $c=ab\backslash a\cup b$ vanishes, though, they contribute to the flux through $\xoverline{AB}$. The analysis here is thus clearer if we refer to the complementary regions instead. From the point of view of the $\xoverline{ab}$ system, these threads contribute to the area, giving the excess ${\rm area}\(\partial \(ab\)\)/4G_N-{\rm area} \(\partial a\cup b\)/4G_N$. Hence these treads are interpreted as classical. However, from the point of view of the $\overline{a\cup b}$ system, these terms contribute to the bulk entropy and hence are interpreted as quantum. In this case, these threads measure the difference in entropies, $S[\xoverline{ab}]-S[\overline{a\cup b}]$, which for pure states is equivalent to $S[a\cup b]-S[ab]$.

As before, a change in the bulk state and, hence, the entropies, can make the connected or disconnected configurations dominate. We will consider these two options next.

\begin{itemize}
  \item \emph{Disconnected solution dominates:} The disconnected solution would dominate if
\bea\label{dc-dom-n}
\frac{{\rm area}\( \partial ab\)}{4G_N}-\frac{{\rm area}\( \partial \(a\cup b\)\)}{4G_N}>S[a\cup b]-S[ab]\,.
\eea
This happens when the disconnected configuration dominates classically, ${\rm area}\( \partial \(a\cup b\)\)<{\rm area}\( \partial ab\)$, at least by the amount given on the right hand side of (\ref{dc-dom-n}).
  \item \emph{Connected solution dominates:} The connected solution would dominate if
\bea\label{cc-dom-n}
\frac{{\rm area}\( \partial ab\)}{4G_N}-\frac{{\rm area}\( \partial \(a\cup b\)\)}{4G_N}<S[a\cup b]-S[ab]\,.
\eea
There are two options to satisfy this condition. First, if the connected configuration dominates classically, ${\rm area}\( \partial ab\)<{\rm area}\( \partial \(a\cup b\)\)$, regardless of the bulk entropy. Additionally, there could also be a situation when the \emph{disconnected} configuration dominates classically, ${\rm area}\( \partial ab\)>{\rm area}\( \partial \(a\cup b\)\)$, provided the difference is bounded by the difference in bulk entropies as stated in (\ref{cc-dom-n}).
\end{itemize}

This last configuration highlights an important aspect of the quantum bit threads program that has not been emphasized until now, but deserves clarification. As discussed around equation (\ref{eq:flowsHI}), we can always decompose a flow in terms of a homogeneous solution $v_\text{h}$ and an inhomogeneous one $v_\text{i}$, which lead to classical and quantum thread bundles, respectively. The program (\ref{eq:flowFLM}) maximizes the flow over a region $A$ but, a priori, it seems that it makes no distinction between the two type of flows or threads. However, this is not entirely the case. The program (\ref{eq:flowFLM}) is only fully specified after a density function $s(x)$ is given, which in turn poses a constraint on the flux of $v_\text{i}$ through $\partial \Sigma$. Namely, once an $s(x)$ is given, we already know how many \emph{quantum} threads will end up in $\partial\Sigma$, and this introduces a constraint on the maximization problem. Thus, it appears that the max flow prescription requires filling in the quantum threads first (taking into account their orientation) and only once all these threads have been included one can then start filling in the classical threads. If done in the opposite order one may end up in a situation where sources or sinks are not utilized, e.g. due to the norm bound, and thus the impossibility of actually computing the RHS of (\ref{eq:flowFLM}). This observation will be important in the next section, where we prove a set of general properties of our quantum bit thread prescription.

\section{General properties of the flow program\label{sec:GeneralProps}}
\subsection{Properties of quantum bit threads}
In this subsection we will show that basic properties of classical bit threads such as nesting and the existence of a max multiflow, can be upgraded to our quantum version with some minimal modifications. The general idea of our proofs is to first reduce our quantum setup into a classical one via a projective mapping, then apply the theorems of \cite{Headrick:2017ucz,Cui:2018dyq} to the effective classical problem and, finally, map the solution back to the original quantum setup. We believe that one could alternatively follow the methodology developed in \cite{Headrick:2017ucz,Cui:2018dyq}  by directly manipulating our convex optimization program and so we expect to report some progress in that direction as well in the future.

\subsubsection{Nesting\label{nesting}}

The nesting property for classical bit threads states that given a Riemannian manifold $\Sigma$ with boundary, and a set of nested boundary regions, say $\{A_1, \cdots, A_n\}$ with $A_i\subset A_{i+1}$, there exists a flow $v^{(0)}$ such that its flux through each of the $A_i$ is maximum, 
\bea
S^{(0)}[A_i]=\int_{A_i}\!v^{(0)} \qquad{\rm for\,\,  all \,\, } i\,.
\eea
This statement was proved in \cite{Freedman:2016zud,Headrick:2017ucz,Cui:2018dyq}. If a similar statement holds for our quantum max flow program, such program would require as a minimal condition to have a density $s(x)$ such that its integrated value over each homology region $a_i$ (associated to each boundary region $A_i$) reproduce their corresponding bulk entropies, i.e.,
\bea\label{dens:const}
S[a_i]=\int_{a_i}s(x) \qquad{\rm for\,\,  all \,\, } i\,.
\eea
Only in this way we could define a program from which we can compute all entropies $S[A_i]$.

Let us first show that, provided that the entropies of all bulk regions involved are known, one could construct such density. Given a set of strictly nested boundary regions $A_i\subset A_{i+1}$ for $i \in \{1, \cdots n-1\}$, the bulk homology regions associated to them (this is to $A_i$ we associated the bulk region $a_i$) inherit the same nested relation, namely $a_i \subset a_{i+1}$ for $i \in \{1, \cdots n-1\}$. This follows from the property of entanglement wedge nesting (EWN). This means that the bulk region $a_n$ contains all other regions, and thus one can further separate that region into $n$ non-overlapping parts in the following way:
\bea
a_n=a_1\cup\(a_2\backslash a_1\)\cup \cdots \(a_{n}\backslash a_{n-1}\)\,.
\eea
Since the nested condition is strict, all the bulk regions appearing in the above separation are non-empty and thus the integrated entanglement density on such regions can acquire any value.
Therefore, we define $s(x)$ to satisfy
\bea
\int_{a_1}\!s(x)=S[a_1]\quad{\rm and}\quad \int_{a_{i+1}\backslash a_i}\!\!s(x)=S[a_{i+1}]-S[a_{i}]\qquad {\rm for \,\, all \,\, }i\in\{1, \cdots , n-1\}\,.
\eea
Any density function satisfying these relations will automatically satisfy (\ref{dens:const}).

There is a caveat about the densities defined above, however. We cannot make the densities arbitrarily large point-wise, as this could lead to a violation of the norm bound.  See the end of section \ref{ConvexOpt}, particularly footnote \ref{footnote-maxdensity} for a discussion on this point. This means that we should spread out the densities so that locally they are at most of order $\mathcal{O}(G^{0}_N)$. In line with this choice we can also separate any flow $v$ as in (\ref{eq:flowsHI}), this is, into homogeneous and inhomogeneous pieces $v=v_{\text{h}}+v_{\text{i}}$, with $v_{\text{i}}\sim \mathcal{O}(G_N)$, and guarantee in this way that any inhomogeneous component $v_{\text{i}}$ will safely obey the norm bound constraint strictly.

Using the above separation into homogeneous and inhomogeneous parts, and considering the discrete version of the flows (as directed threads of Planckian thickness), we could now think of constructing the inhomogeneous component of the thread configuration (i.e., the quantum threads) such that they satisfy the constraint equations for the nested set of bulk regions of interest $\{a_i\}$. We can be sure that such thread configuration can be used as part of the solution to the max flow problem for each of the nested regions. In order to guarantee that, we simply need to impose that the quantum threads that end at the sinks in $a_i$  to start at their associated boundary region $A_i$, while those that are sourced in $a_i$ to end at any point in $\bar{A}_n$ (the complement of the largest region).\footnote{Any bulk-to-bulk thread can be turn into the sum of a classical thread and a two quantum threads via the following procedure: First, we elongate each side of the thread (while keeping their endpoints fixed) until it touches the boundary. Such a thread can then be interpreted as sum of two quantum threads (which connect a boundary point with a bulk point) plus a classical thread (which connects two boundary points). The classical thread is then added to the homogeneous piece.\label{footnote-thread-separation}} This last condition is imposed so that no negatively sourced thread contribute to the flux of any of the nested boundary regions $\{A_i\}$. Additionally we restrict all the threads defined above to cross at most once any extremal surface associated to the set of nested boundary regions. After the above choices are made, then, the number of quantum threads connecting each boundary region $A_i$ with $a_i$ will provide the maximum quantum contribution of the flux through $A_i$ for all the boundary regions while maximally avoiding the various extremal surfaces.

After all the previous work we can now reduce the remanent part of the flow maximization problem (i.e., filling in the classical threads) by a collapsing procedure in which all the quantum threads together with the space occupied by them is collapsed to zero size. We can think of this step as shrinking the quantum threads to zero thickness.  This removes the quantum threads from the problem, though, leaving a trace through curves of conical singularities along the location of the threads \footnote{We should further impose some length bound prescription on the previous construction to avoid arbitrarily large threads. Doing that would avoid the possibility of having a resulting geometry (after the collapsing process) modified macroscopically and thus making the validity of the classical flow theorems questionable.}. The resulting effective manifold $\Sigma'$, therefore, does not obey strong continuity properties. Nevertheless, the max flow theorems coming from network theory are robust and can be applied in such manifolds. Applying the classical nesting property on $\Sigma'$ for the resulting set of nested boundary regions $\{A'_i\}$ will then guarantee the existence of a flow whose flux through every $A_i'$ is maximal. Once one such flow is constructed we can recover the original manifold together with the quantum threads by reverting the collapsing process, and thus we end up with a set of classical and quantum nested threads with maximum flux through $A_i$ in the original manifold $\Sigma$.\footnote{Notice that the classical threads constructed for $\Sigma'$ could cross any of the many singular curves (curves of conical singularities where the quantum threads were originally located). One can remedy this via an infinitesimal transformation which minimally move the threads so that to avoid all those crossings.}

To summarize, then, we have shown that a max flow program with entanglement density $s(x)$ obeying the set of relations
\bea
S[a_i]=\int_{a_i}s \qquad{\rm for\,\,  all \,\, } i\,,
\eea
where $a_i$ are the homology regions associated to a set of nested boundary regions $\{A_i\}$ (i.e., $A_i\subset A_{i+1}$), have always a max flow solution $v$ such that
\bea
S[A_i]=\int_{A_i}v \qquad{\rm for\,\,  all \,\, } i\,.
\eea
This is the quantum version of the nesting property for classical bit threads.

\subsubsection{Max multiflow \label{max-mf}}

For classical bit threads, the definition of a multiflow and a theorem that states the existence of a max multiflow, was introduced and proved in  \cite{Cui:2018dyq}. We will review those here and describe how to use these results together with the collapsing process of the quantum bit threads introduced in the previous subsection to prove the equivalent statements for our quantum max flow program.

We will start with the definition of a classical multiflow. Given a Riemanninan manifold $\Sigma$, with boundary $\partial \Sigma$, let $A_1$, $\cdots ,A_n$ be non-overlapping regions of $\partial \Sigma$ covering $\partial \Sigma$ $\(\cup_i A_i=\partial \Sigma\)$. A multiflow is a set of vector fields $v_{ij}$ on $\Sigma$ satisfying the following conditions:
\bea
&&v_{ij}=-v_{ji}\,, \qquad \hat{n}_k \cdot v_{ij}=0 \quad {\rm on}\quad A_k \quad \(k\neq \{i,j\}\)\,, \label{mmf-1-a}\\ \label{mmf-1-b}
&& \nabla \cdot v_{ij}=0\qquad {\rm and}\qquad \sum_{i<j}^n|v_{ij}|\leq \frac1{4G_N}\,.
\eea
From these defining properties one has that, given a set of coefficients $-1<\xi_{ij}<1$, then
\bea\label{multiflow}
v\equiv \sum_{i<j}^{n} \xi_{ij}v_{ij}
\eea
is a flow, i.e., $v$ is divergenceless and norm bounded. Given a multiflow $\{v_{ij}\}$ we can define a collection of $n$ vector fields according to
\bea
v_i\equiv\sum_{j=1}^nv_{ij}\,,
\eea
whose flux on $A_i$ is bounded by its entropy
\bea\label{maxmultiineq}
\int_{A_i}v_i\leq S^{(0)}[A_i]\,.
\eea
The \emph{max multiflow theorem} states that there exist a max multiflow $\{v^*_{ij}\}$ such that for each $i$, $v^*_i$ is a max flow for $A_i$. This is, for such a flow all inequalities in (\ref{maxmultiineq}) are strictly saturated,
\bea
\int_{A_i}v^*_i=S^{(0)}[A_i]\,.
\eea

Note that (\ref{multiflow}) implies that \emph{any} flow $v$ constructed from adding all the independent multiflow components $\{v_{ij}\}$ with coefficients $\xi_{ij} \in \{-1,1\}$ is a consistent flow. However, in our quantum generalization this would not be possible, because changing the orientation of a given quantum bundle would change the sign of the entanglement density sourcing that bundle, and thus change the flow program. This can be traced down to the fact that the divergence constraint (\ref{eq:constr}) is not invariant under parity $v\to-v$ in the presence of sources, and implies that in order to properly describe a multiflow in the quantum case, one would need to include a set of bundle densities as well.

In view of this last observation, we propose the following generalizations of the definition of a multiflow and the max multiflow theorem:
Given a Riemanninan manifold $\Sigma$, with boundary $\partial \Sigma$, let $A_1$, $\cdots ,A_n$ be non-overlapping regions of $\partial \Sigma$ covering $\partial \Sigma$ $\(\cup_i A_i=\partial \Sigma\)$. A \emph{quantum multiflow} is a set of vector fields $v_{ij}$ and associated sources $s_{ij}(x)$ on $\Sigma$ satisfying the following conditions:
\bea\label{q-bundle-1}
&&v_{ij}=-v_{ji}\,, \qquad \hat{n}_k \cdot v_{ij}=0 \quad {\rm on}\quad A_k \quad \(k\neq \{i,j\}\)\,, \label{mmf-1-a}\\ \label{mmf-1-b}\label{q-bundle-2}
&& \nabla \cdot v_{ij}=-4G_N s_{ij}(x)\qquad {\rm and}\qquad \sum_{i<j}^n|v_{ij}|\leq \frac1{4G_N}\,.
\eea
The entanglement densities $s_{ij}(x)$ satisfy the properties
\bea\label{q-s-bundle}
s_{ij}(x)=-s_{ji}(x)\,, \quad{\rm and }\quad s_i(x)\equiv \sum_{j=1}^n s_{ij}(x) \,,
\eea
where the first condition is inherited from the antisymmetry of the flow bundle (\ref{q-bundle-1}),
and the set of densities $\{s_i(x)\}$ are required to represent entanglement densities for the homology regions $\{a_i\}$, i.e.,
\bea\label{ent-density-si}
\int_{a_i}\!s_i(x)=S[a_i]\,. \qquad {\rm for  \quad }i=\{1, \cdots , n\}\,.
\eea
Given a quantum multiflow, we can now define a set of $n$ vector fields according to
\bea\label{vec-of-flows}
v_i\equiv \sum_{j=1}^nv_{ij}\,,
\eea
whose fluxes across their corresponding regions are bounded by their entropies,
\bea\label{Qmaxmultiineq}
\int_{A_i}\!v_i\leq S[A_i]\,.
\eea
In other words the flows $v_i$ obey the right divergence constraint on $a_i$ which is guaranteed by condition (\ref{ent-density-si}) and the definition in (\ref{q-s-bundle})

We are now ready to state the quantum version of the max multiflow theorem, which we denote as \emph{quantum max multiflow theorem}: There exists a max multiflow $\{v^*_{ij}\}$ with associated sources $\{s^*_{ij}(x)\}$ such that for each $i$, $v^*_i$ is a max flow on $A_i$. In particular this means all the sources has been utilized by the quantum component of the thread bundle

This is, for such a flow all inequalities in (\ref{Qmaxmultiineq}) must be strictly saturated,
\bea
\int_{A_i}\!v^*_i=S[A_i]\,.
\eea
To prove this theorem we will follow the same strategy used to prove nesting. We start by separating each component of a multiflow $\{v_{ij}\}$ into two pieces, $v_{ij}=v^\text{h}_{ij}+v^{\text{i}}_{ij}$, where superscripts `$\text{h}$' and `$\text{i}$' stand for the homogeneous and inhomogeneous parts, respectively (or, equivalently, classical and quantum parts). We take the bundle $\{v^{\text{i}}_{ij}\}$ to be a \emph{minimal} quantum multiflow, this is a multiflow which include only quantum threads that connect points in the bulk with points in the boundary (see footnote \ref{footnote-thread-separation} for a comment on this point) while satisfying the constraint equations (\ref{mmf-1-a})-(\ref{mmf-1-b}) and crossing at most once any extremal surface associated to the boundary regions $A_i$. Notice that the flow $\{v^{\text{i}}_{ij}\}$ constructed in this way will contain the information of the bundle density component $s_{ij}(x)$, which is carried by the bulk endpoints of the associated threads, taking into account their orientation. A bundle density will then give us detailed information about the endpoints of the associated minimal flow bundle $\{v^{\text{i}}_{ij}\}$.

Using only this kind of threads one can show that the set of $n$ flows defined as
\bea
v_i^\text{i}=\sum_{j=1}^n v^{\text{i}}_{ij}
\eea
can be chosen such that they contribute to the flux on the regions $A_i$ maximally while maximally avoiding the min cuts.

To show one construction of a particular density bundle $\{s_{ij}(x)\}$ (and therefore of their associated minimal quantum flow bundle $\{v_{ij}^\text{i}\}$), we start by defining an auxiliary density $\tilde{s}(x)\equiv\sum_i\tilde{s}_i(x)\chi_{a_i}(x)$ with $\tilde{s}_i(x)$ obeying
\bea
\int_{a_i}\!\tilde{s}_i(x)=S[a_i]\,,
\eea
where $\chi_{a_i}(x)=1$ on $a_i$ and $0$ everywhere else. Notice that this density is defined globally.

Using $\tilde{s}(x)$ we can now construct a global \emph{minimal} quantum flow $v^\text{i}$ such that their associated threads connect points in the boundary region $A_i$ with all the sinks (of positive entanglement density) in $a_i$, and connect all the sources (of negative entanglement density) in $a_i$ to points in the complementary boundary region $\bar{A}_i$. Note that this last property also imply that there will generically be treads that connect points on $\cup_{j\neq i}a_j$ with points on $A_i$. These last threads contribute negatively to the flux on $A_i$ and therefore, the flux through $A_i$ of the constructed flow $\tilde{v}^\text{i}$ will not be maximal. Nevertheless, we will show how taking this thread configuration as our starting point, we can build a set of flows $v_i^{\text{i}}$ (and associated densities $\{s_i\}$ ) with maximal fluxes through their associated boundary regions $A_i$.

We now proceed as follows. First, we build a set of $n$ entropy densities defined as
\bea
s_i(x)\equiv\tilde{s}_i(x)\chi_{a_i}(x)-\sum_{j\neq i} \tilde{s}_j(x)\chi_{a_j}(x)\,.
\eea
The resulting densities $s_i(x)$ will still represent a consistent entanglement density for the associated homology region $a_i$.
For each $s_i(x)$ we can build an associated flow $v^\text{i}_i$ from the $v^i$ constructed previously by reverting the signs of the threads that start or end at $\cup_{j\neq i}a_j$, as a result, the fluxes through $A_i$ of the $v^\text{i}_i$ are now maximal. To obtain the full quantum bundle we would need an extra 
 separation of the flows  $v_i^{\text{i}}$ into components with the appropriate imposed symmetry, $v^{\text{i}}_{ij}=-v^{\text{i}}_{ji}$, which result in a similar separation for $s_i(x)$ into $s_{ij}(x)=-s_{ji}(x)$ by the constraint (\ref{q-bundle-2}). This leads to the full construction of the quantum component of a max multiflow $\{v_{ij}^{\text{i}}(x)\}$, $\{s_{ij}(x)\}$.\footnote{This separation is always possible because we have constructed the set of $\{v_i^{\text{i}}\}$ and $\{s_i(x)\}$ by simply reversing the signs of the densities and flow bundles in some regions of a single $v$ and $\tilde{s}$.}

The final step is to obtain the classical bundle component $\{v_{ij}^{\text{h}}(x)\}$. To do this, we carry out the same collapsing procedure described in subsection (\ref{nesting}). We then apply the classical max multiflow theorem, which guarantees the existence a max $\{v_{ij}^{\text{h}}\}$  on the resulting manifold $\Sigma'$. Finally, reversing the collapsing  process, we bring back the quantum component of the max multiflow $\{v_{ij}^{\text{i}}\}$ together with their associated multi density $\{s_{ij}(x)\}$. In combination, then, they provide a full solution to quantum max multiflow problem in the original manifold $\Sigma$. This concludes our proof of the quantum max multiflow theorem.

\subsection{Properties of holographic entanglement}
The entanglement entropy associated to spatial regions of any quantum field theory is known to obey certain general entanglement relations. Among the most fundamental ones are the subadditivity and strong subadditivity inequalities. In this subsection, we will show that our max flow program
imply both subadditivity and strong subadditivity of boundary entanglement entropies, thus, providing in this way a consistency check of our proposal.

\subsubsection{Subadditivity}
Subadditivity states that given two non-overlapping boundary regions $A$ and $B$ ($A\cap B=\emptyset$), the following relation between the associated entanglement entropies must be satisfied:
\bea\label{SA-1}
S[A]+S[B]\geq S[A\cup B]\,.
\eea
Thus, we have three boundary regions $A$, $B$ and $A\cup B$ and three associated bulk homology regions\footnote{Assuming a unique homology region for each boundary region.} which we will denote $a$, $b$ and $ab$, respectively. Notice that, in general, $ab\neq a\cup b$. EWN then implies the following relations. First, since $A\subseteq A\cup B$ and $B\subseteq A\cup B$, then $a\subseteq ab$, $b\subseteq ab$ and thus $a\cup b \subseteq ab$. Similarly, $A\cap B=\emptyset$ implies $a\cap b=\emptyset$.

We are now interested in proving the subadditivity property of boundary entropies. In order to do so we will assume subadditivity of the bulk entropies, which should be true for any consistent quantum theory. We will also use the above nesting relations between bulk regions and pick a physically motivated entanglement density $s(x)$, such that it allows us to compute a set of relevant entanglement entropies using the \emph{same} flow program.

We begin with the subadditivity inequality for bulk entropies, which takes the form
\bea
\label{bulk-SA}
S[a]+S[b]\geq S[a\cup b]\,.
\eea
The left hand side quantities in the boundary inequality (\ref{SA-1}) are computed by the QES formula or its flow equivalent. This suggests the following defining constraints on the density:
\bea\label{const-s-SA}
\int_{a}s(x)=S[a] \qquad {\rm and}\qquad \int_{b}s(x)=S[b]\,,
\eea
so that we have a well defined max flow program for the individual regions $A$ and $B$. On the other hand, the right hand side of (\ref{SA-1}) involves an extra, possibly non-empty region, $ab\backslash a\cup b$. The density there is not affected by the above constraints. A consistent choice for the value of the integrated density on $ab\backslash a\cup b$ should however, vanish when $ab=a\cup b$. In principle, a density satisfying only (\ref{const-s-SA}) could represent a valid entanglement density even for $ab$ in some particular situations. For instance, if $ab\backslash a\cup b=\emptyset$ and subadditivity for $a$ and $b$ is strictly saturated, then $\int_{a\cup b}s(x)=S[a\cup b]$ so $s(x)$ will be a valid entanglement density for $a$, $b$, and also $ab$. More generally, we can impose the following constraint:
\bea\label{const-s-SA-6}
\int_{ab\backslash a\cup b\neq\emptyset}s(x)= S[ab]-S[a\cup b]\,,
\eea
so that $s(x)$ would represent an entanglement density also for $ab$, in cases where  subadditivity for $a$ and $b$ is strictly saturated and $ab\backslash a\cup b\neq\emptyset$. Note that this last constraint is also self-consistent, in the sense that its LHS and RHS vanish exactly when $ab=a\cup b$. Interestingly, any density satisfying (\ref{const-s-SA}) and (\ref{const-s-SA-6}) will imply subadditivity for the boundary theory. We will now show this by explicit calculation.

Since any density $s(x)$ that obeys (\ref{const-s-SA}) and (\ref{const-s-SA-6}) defines a convex program $\mathcal{F}_{AB}$, any flow $v\in\mathcal{F}_{AB}$ immediately satisfies the following inequality:
\bea
S[A]+S[B]\geq \int_A v +\int_B v=\int_{A\cup B} v\,.
\eea
Taking a flow of this program with maximum flux through $A\cup B$ and using the max flow-min cut theorem we obtain:
\bea\label{SA-2}
S[A]+S[B]\geq \max_{v\in \mathcal{F}_{AB}}\int_{A\cup B} v=\frac{{\rm Area}[\tilde{\partial} ab]}{4G_N}+\int_{ab} s(x)\,,
\eea
where we have introduced the notation $\tilde{\partial }ab\equiv\partial ab\backslash \partial \Sigma$.
The density integral can be separated into various bulk regions. Using the constraints (\ref{const-s-SA}) and (\ref{const-s-SA-6}) we obtain,
\bea\label{const-s-SA-3}
\int_{ab} s(x)&=&\int_{a\cup b}s(x)+\int_{ab\backslash a\cup b}\!\!s(x)=S[a]+S[b]+S[ab]-S[a\cup b]\,.
\eea
Hence, subadditivity of bulk entropy (\ref{bulk-SA}) implies
\bea\label{const-s-SA-5}
\int_{ab} s(x)\geq S[ab]\,.
\eea
Plugging this into (\ref{SA-2}) leads to the subadditivity inequality of boundary entropies (\ref{SA-1}), and this concludes our proof.

\subsubsection{Strong subadditivity}
Strong subadditivity states that given three non-overlapping regions $A$, $B$ and $C$ ($A\cap B=\emptyset,$ $ A\cap C=\emptyset$ and $B\cap C=\emptyset$) the following relation between the associated entanglement entropies must be satisfied:
\bea\label{SSA}
S[A\cup B]+S[B\cup C]\geq S[B]+S[A\cup B\cup C]\,.
\eea
The above statement can be equivalently written as
\bea\label{SSA-2}
S[A_1]+S[A_2]\geq S[A_1\cap A_2]+S[A_1\cup A_2]\,,
\eea
with the following identifications
\bea\label{Ai-1}
A_1=A\cup B, \quad A_2=B\cup C\,.
\eea
We are now interested in proving the strong subadditivity inequality for boundary entropies (\ref{SSA}). We will follow the same logic as for subadditivity. Namely,  we will assume that strong subadditivity for bulk entropies hold (which must be true for any consistent quantum theory), the nesting relations between bulk regions and a physically motivated choice of entanglement density $s(x)$ such that a set of the relevant entropies can be computed from a unique flow program.  As before, we start with strong subadditivity of the bulk regions of interest:
\bea\label{SSA-2.5}
S[ab]+S[bc]\geq S[ab\cap bc]+S[ab\cup bc]\,.
\eea
which follows from (\ref{SSA-2}) with the identifications $A_1=ab$, $A_2=bc\,$, where $ab$ and $bc$ are the homology regions associated to $A\cup B$ and $B\cup C$, respectively. Apart from the bulk regions $ab$ and $bc$ involved in (\ref{SSA-2.5}), the boundary equation (\ref{SSA}) involves various other bulk regions, which we will define below. The idea here is to find a minimum set of constraints of $s(x)$ over these extra regions and define a flow program that can be used to compute the relevant entropies entering in the strong subadditivity inequality of boundary entropies.

Let us begin by studying the implications of EWN. First, notice that $A\subseteq A\cup B$ implies $a\subseteq ab$; likewise $B\subseteq A\cup B$ and  $B\subseteq B\cup C$ imply $b\subseteq ab$ and $b\subseteq bc$, respectively. These last two relations combined imply $b\subseteq ab\cap bc$. Similarly,  $A\cup B\subseteq A\cup B\cup C$ and $B\cup C\subseteq A\cup B\cup C$ imply
$ab\subseteq abc$ and $bc\subseteq abc$, which in turns leads to
$ab\cup bc \subseteq abc$. Some of the above inclusion relations are strict provided that the individual boundary regions $A$, $B$ and $C$ are non-empty; others, on the other hand, can become equalities. This distinction will be relevant a bit later.

Next, we will exploit the above nested relations to divide various bulk regions into non-overlapping parts.  To do so, we start from the largest bulk region that we can define based on the boundary regions $A$, $B$ and $C$. This region is $abc$ as implied by EWN. Then, we proceed by dividing this bulk region into four parts, according to
\bea\label{abc-separation}
abc=\(abc\backslash ab\cup bc\) \cup \(ab\backslash bc\) \cup \(bc \backslash ab\)\cup \(ab \cap bc\)\,.
\eea
Similarly, we separate the bulk regions $ab$ and $bc$ into two parts each,
\bea\label{ab-bc-partitions}
ab&=& \(ab\backslash bc \)\cup \(ab \cap bc \)\,, \nonumber \\
bc&=& \(bc \backslash ab \)\cup \(ab \cap bc\)\,.
\eea
One of them, $ab \cap bc$, is present in both, so we will have at least three non-overlapping bulk regions. When necessary, we can further divide this intersection into two parts as
\bea
ab \cap bc=\[(ab \cap bc )\backslash b\] \cup b\,.
\eea
We point out that the partitions of the above regions are possible given the inclusion relations previously discussed. However, some of these regions could be empty. Assuming that the individual boundary regions $A$, $B$ and $C$ are non-empty, the only regions involved in the above partitions which could be empty are
\bea \label{empty-sets?}
\(abc\backslash ab\cup bc\)  \qquad {\rm and} \qquad \[(ab \cap bc )\backslash b\]\,.
\eea

Now, recall that the boundary entropies appearing on the LHS of (\ref{SSA}) include the entropies associated to the bulk regions $ab$ and $bc$. This motivates the following constraints on the density $s(x)$,
\bea \label{ent-density-SSA}
\int_{ab}s(x)=S[ab] \qquad {\rm and }\qquad \int_{bc}s(x)=S[bc]\,,
\eea
so that we have a well defined max flow program for these regions. On the other hand, note that from the decompositions in (\ref{ab-bc-partitions}) it follows that any entanglement density $s(x)$ defined on those regions should satisfy the following equation:
\bea\label{SSA-3}
\int_{ab}s(x)+\int_{bc}s(x)=\int_{ab\cap bc}\!\!s(x)+\int_{ab\cup bc}\!\!s(x)\,.
\eea
This is true as each side of the equation covers the same regions the same number of times. Now, bulk strong subadditivity (\ref{SSA-2.5}) together with this last equation imply the following constraint on the integrated density
\bea\label{SSA-4}
\int_{ab\cap bc}s(x)+\int_{ab\cup bc}s(x)\geq S[ab\cap bc]+S[ab\cup bc]\,.
\eea
Next, we need to compare the bulk entropies in the RHS of (\ref{SSA-4}) with the bulk entropies appearing on the RHS of the boundary strong subadditivity (\ref{SSA}), i.e., $S[b]+S[abc]$, to determine what bulk regions we have left out. To do this, we simply rewrite the integrated densities over these regions as
\bea\label{decomp-abc-b}
\int_{b}s(x)&=&\int_{ab\cap bc}\!\!s(x)-\int_{\(ab\cap bc\)\backslash b}\!\!s(x)\,, \\
\int_{abc}\!\!s(x)&=&\int_{ab\cup bc}\!\!s(x)+ \int_{abc\backslash \(ab\cup bc\)}\!\! s(x)\,.\label{decomp-abc-b2}
\eea
Thus we have generically two extra, non-empty regions whose integrated densities have not been constrained so far. These regions are precisely the ones which can be empty in some configurations, indicated in (\ref{empty-sets?}). A natural choice here is to impose the following constraints:
 \bea\label{SSA-5}
 \int_{\(ab\cap bc\)\backslash b}\!\!s(x)&=& S[ab\cap bc]-S[b]\,, \\
 \label{SSA-6}
 \int_{abc\backslash \(ab\cup bc\)}\!\! s(x)&=&S[abc]-S[ab\cup bc]\,,
 \eea
which can be deduced following the same logic as in the previous section. Notice that these constraints are self-consistent. Namely, both sides of the two equations vanish whenever $ab\cap bc=b$ and $abc=ab\cup bc$, respectively.


With the above choices, i.e., for a density that satisfies (\ref{ent-density-SSA}), (\ref{SSA-5}) and (\ref{SSA-6}) we now have a well defined program that can be used to test the boundary SSA inequality (\ref{SSA}). Note that any entanglement density that satisfies such constraints defines a convex program $\mathcal{F}_{ABC}$, and  any flow in that program $v\in\mathcal{F}_{ABC}$ will satisfy
\bea
S[X]\geq \int_{X} v\,,
\eea
where $X\in\{A\cup B, B\cup C\}$. As usual, the equality is saturated by maximizing the flux through the corresponding region, i.e.,
\be
S[X]= \max_{v\in \mathcal{F}_{ABC}}\int_{X} v\,.
\ee
Now, for a flow $v$ in this program we have that
 \bea
 S[A\cup B]+S[B\cup C]\geq \int_{A\cup B} v +\int_{B\cup C} v= \int_{A\cup B\cup C} v +\int_{B} v\,,
 \eea
 where in the last equation we have just rearranged the boundary regions. Next, we chose a flow $v$ which is a max flow  for the boundary regions $B$ and $A\cup B\cup C$ simultaneously. This is guaranteed to exist by the nesting property of max flows (see section \ref{nesting}), and leads to
  \bea
 S[A\cup B]+S[B\cup C] \geq \max_{v \in \mathcal{F}_{ABC}}\int_{A\cup B\cup C}\!\! v +\max_{v \in \mathcal{F}_{ABC}}\int_{B} v\,.
 \eea
 Using the max flow-min cut theorem, the RHS of the above equation yields
 \bea\label{max-flow-AB-BC}
 \max_{v \in \mathcal{F}_{ABC}}\int_{A\cup B\cup C}\!\! v+\max_{v\in \mathcal{F}_{ABC}}\int_{B} v=\frac{{\rm Area}[\tilde{\partial}abc]}{4G_N}+\int_{abc}\!s(x)+\frac{{\rm Area}[\tilde{\partial} b]}{4G_N}+\int_{b}s(x)\,.
 \eea
 The integrated densities appearing on the RHS of the above equation are fixed by the constraints (\ref{ent-density-SSA}), (\ref{SSA-5}) and (\ref{SSA-6}). Decomposing each integral as in (\ref{decomp-abc-b})-(\ref{decomp-abc-b2}) and using (\ref{SSA-3}) it then follows that
 \bea\label{SSA-7}
 \int_{abc}\!s(x)+ \int_{b}s(x)=S[abc]+S[b]+S[ab]+S[bc]-S[ab\cup bc]-S[ab\cap bc]\,.
 \eea
Bulk strong subadditivity implies that the sum of the last four terms in (\ref{SSA-7}) is non-negative and thus we conclude that
 \bea
  \int_{abc}\!s(x)+ \int_{b}s(x) \geq S[abc]+S[b]\,.
 \eea
Finally, plugging this inequality into (\ref{max-flow-AB-BC}) leads to the strong subadditivity inequality for the boundary entropies (\ref{SSA}). This concludes our proof.

\subsection{The fate of holographic monogamy}
In section \ref{max-mf} we presented a quantum max multiflow theorem and its proof. Akin to the classical case, in which the monogamy of mutual information can be seen as a straightforward consequence of the max multiflow theorem, it is natural to ask whether this is also the case for the quantum version. Of course, it is not expected that the monogamy will hold at $\cO(G^0_N)$ generically, since it is only at infinite $c$ that such a property can be derived from geometric arguments. However, it should be interesting to understand if there are special conditions on the bulk entropies that would imply such a property at next-to-leading order, as this could provide hints to discriminate between CFTs with potential semi-classical duals. This question becomes relevant in scenarios of double holography, where the corrections to the entanglement entropy formula come from a similar area-like term in the second holographic layer, thus obeying the monogamy property. Indeed, in this section we will show that for a certain natural choice of entanglement densities we can derive the monogamy property at $\cO(G^0_N)$, provided one imposes an extra condition on the max multiflow bundle of entropy densities $\{s_{ij}(x)\}$. Such a feature is in line with their interpretation as bulk EPR-like entanglement and it can be shown to be equivalent to requiring the monogamy of bulk entropies.

Following the notation of section \ref{max-mf} we separate the full bit thread bundle into its homogeneous and inhomogeneous components $\{v_{ij}\}=\{v^{\rm h}_{ij} \} + \{ v^{\rm i}_{ij}\}$, with the additional requirement of minimality on $ \{ v^{\rm i}_{ij}\}$. In analogy with the classical counterpart, the $\{v^{\rm h}_{ij}\}$ component is known to imply the monogamy property on the effective manifold obtained after the collapsing process has been carried out on the minimal $\{v^{\rm i}_{ij}\}$ component, following the procedure of \cite{Cui:2018dyq}. However, there is no similar statement for the quantum component $ \{v^{\rm i}_{ij}\}$, in other words, carrying out the same analysis to this component does not imply an analogue for bulk entropy monogamy. The issue boils down to the fact that $\{v^{\rm i}_{ij} \}$ does not obey a geometric constraint as the $\{v^{\rm h}_{ij} \}$ does. We will derive under which conditions the full $\{v_{ij}\}$ is such that the boundary entropies obey the monogamy property.

We will follow the strategy of \cite{Cui:2018dyq} for the derivation of the classical MMI. Given a max multiflow bundle $\{v_{ij}\}$ on $\Sigma$ such that $\partial \Sigma$ is separated into four regions $\{A, B, C, D\}$ we construct the following flows:
\bea\label{u-flows}
u_1&=&v_{AC}+v_{AD}+v_{ BC}+v_{BD}=\frac 12\(v_A+v_B-v_C-v_D\)\,,  \nonumber  \\
u_2&=&v_{AB}+v_{AD}+v_{ CB}+v_{CD}=\frac 12\(v_A-v_B+v_C-v_D\) \,, \nonumber \\
u_3&=&v_{BA}+v_{BD}+v_{ CA}+v_{CD}=\frac 12\(-v_A+v_B+v_C-v_D\)\,.
\eea
Each of these flows will have an associated density $\bar{s}_i(x)$ built from the bundle of densities $\{s_{ij}\}$ associated to the flow bundle $\{v_{ij}\}$ which appears on the quantum max multi-flow construction. Essentially, we superpose the sources as dictated by the combinations appearing in (\ref{u-flows}). Writing down those densities explicitly, we obtain:
\bea\label{bar-s}
\bar{s}_1(x)&=&s_{ac}(x)+s_{ad}(x)+s_{bc}(x)+s_{bd}(x)\,, \nonumber \\
\bar{s}_2(x)&=&s_{ab}(x)+s_{ad}(x)+s_{cb}(x)+s_{cd}(x)\,, \nonumber \\
\bar{s}_3(x)&=&s_{ba}(x)+s_{bd}(x)+s_{ca }(x)+s_{cd}(x)\,.
\eea
Following the procedure outlined in \cite{Cui:2018dyq} we would need to compare the flux of $u_1$ through region $AB$ with $S[AB]$. For that comparison to be useful, one should be able to embed the flow $u_1$ in a manifold with an entanglement density for both $ab$ and $a\cup b$ (a priori the density $\bar{s}_1(x)$ does not satisfy that property). The embedding property is required so that $S[AB]$ can be obtained as a max flow in such a manifold with the prescribed density. Inspired by the density constructions from the previous sections, we construct an $s_1$ on $\Sigma$ that obeys
\bea\label{ent-u1}
\int_{ab\backslash a\cup b}s_1=S[ab]-S[a\cup b] \qquad {\rm and}\qquad \int_{a\cup b}s_1=S[a\cup b]\,,
\eea
so that $s_1$ is an entanglement density for $a\cup b$ and $ab$ simultaneously. We can additionally define $s_1$ to agree with $\bar{s}_1$ on the complement of $ab$, $\(ab\)^c$. Below, we will use $\bar{s}_1$ as an explicit example, though, everything can be straightforwardly generalized to $\bar{s}_2$ and $\bar{s}_3$.

We will impose the following constraint on $\bar{s}_1$:
\bea\label{embedding-condition}
\int_{a\cup b}\bar{s}_1\leq S[a\cup b] \qquad {\rm and}\qquad
 \int_{ab\backslash a\cup b}\bar{s}_1\leq S[ab]-S[a\cup b]\,.
\eea
Notice that if this is the case, then one can embed the flow $u_1$ into the density $s_1$ in the following way. First, we identify the sources described by $\bar{s}_1$ with the ones in $s_1$ leaving behind some net positive sources in the regions $a\cup b$ and $ab\backslash a\cup b$ as allowed by (\ref{embedding-condition}). Since the defining property of $s_1$ only fix the integrated values on those regions (\ref{ent-u1}), we can chose the remaining unpaired sources to be strictly positive.
We can always do this embedding no matter what the classical threads in $u_1$ do (in particular, $u_1$ could saturate the norm bound at the associated min-cut). This is so because one can always increase the flux of a given $u_1$ through the region $AB$ by adding the  extra quantum threads needed to pair with the remaining positive sources present in $s_1$ inside the associated homology region without affecting the classical configuration. As a result, we have the inequality
\bea
\int_{AB}u_1 \leq \max_{v \in \mathcal{F}_{AB}}\int_{AB}v=S[AB]\,,
\eea
valid for any $u_1$ which is the result we were after.\footnote{\label{footnote-16}The potential difficulties present when trying to embed $u_1$ in $s_1$ arise only when the threads in $u_1$ are close to saturating the min-cut associated to $AB$, otherwise, we can always do the embedding. In other words we can always add the needed quantum threads without the risk of saturating the min-cut. This situation will only happen in the vicinity of a phase transition. Away from these situations our quantum max-multiflow would imply MMI without the need of constraints (\ref{embedding-condition}) and their analogues for $\bar{s}_2$ and $\bar{s}_3$} Although we needed to embed $u_1$ into $s_1$ for the above inequality to hold, once we have this result we can think of $u_1$ as being a flow associated with the pair $\(\Sigma, \bar{s}_1\)$. This will be essential in the next constructions since in this way the flows come together with their sources and can be superposed.

Now, let us explore what are the limitations put by the constraints in (\ref{embedding-condition}).
From the explicit construction of $\{s_{ij}\}$ outlined in section \ref{max-mf}, one finds that the bundle of densities have support on $\cup_i a_i$ and, therefore, the same is true for the $\{\bar{s}_i\}$. This means that, in such cases, the left hand side of the second inequality in (\ref{embedding-condition}) equals zero and thus one would be limited to situations in which
\bea\label{ineq-ab-acupb}
S[ab]\geq S[a\cup b]\,,
\eea
for the above embedding to hold.\footnote{This condition is only relevant if the dominant homology region is $ab\neq a\cup b$.} We will assume this since that construction of the $\{s_{ij}\}$ will be our prototypical example. Additionally, we can assume that each $s_{ij}$ has support only on the region $a_i \cup a_j$.

The first inequality in (\ref{embedding-condition}) also puts an important constraint, which we will now analyze in detail. First, from the defining properties of the $\{s_{ij}\}$ we have
\bea\label{S-ents}
S[a]&=&\int_{a}\(s_{ab}+s_{ac}+s_{ad}\)\,, \nonumber \\
S[b]&=&\int_{b}\(s_{ba}+s_{bc}+s_{bd}\)\,, \nonumber \\
S[c]&=&\int_{c}\(s_{ca}+s_{cb}+s_{cd}\)\,.
\eea
Plugging (\ref{bar-s}) into (\ref{embedding-condition}) and using the above relations we get
\bea\label{bulk-mmi-inequality}
S[a]+S[b]-\int_a s_{ab}-\int_b s_{ba}\leq S[a\cup b]\,,
\eea
which can be rewritten as
\bea\label{bound-mutual-mmi}
I[a:b]\leq \int_a s_{ab}+\int_b s_{ba}\,.
\eea
This is an interesting constraint. It presents a bound on the bulk mutual information between two regions in terms of the components of the density bundle that contributes to the bulk entropies of only those regions. In fact, this inequality is not new in the bit thread literature. For classical bit threads, the classical max multiflow theorem implies a similar bound for the boundary mutual information in terms of the flux of the flow components connecting the associated boundary regions:
\bea
I[A:B]\leq \int_A v_{AB}+\int_{B}v_{BA}\,.
\eea
The analogy is more explicit if one thinks of the bulk entropy densities as coming from an emergent geometric setup, in other words, if one thinks of it as in a double holographic scenario. In that case the entropy densities corresponds to the endpoints of some thread configuration that extend in the extra dimension and thus obey the max multiflow theorem which imply the existence of the bundle of entropy densities $\{s_{ij}\}$. Since these densities come from a geometric setup, then, the bulk entropy of the combined region $S[a\cup b]$ will bound the integrated density of any configuration (from the double holography perspective it bounds the flux on any flow through the region $a\cup b$). In particular, for the one provided by the max multiflow construction, this implies
\bea
S[a\cup b]\geq \int_{a\cup b}\(s_{ac}+s_{ad}+s_{bc}+s_{bd}\)\,,
\eea
which is the same as (\ref{bulk-mmi-inequality}) using (\ref{S-ents}). Thus the bound (\ref{bound-mutual-mmi}) would be guaranteed provided the bulk entropy densities can be obtained geometrically.

Generalizing the above analysis for $u_2$ and $u_3$, and assuming the relevant homology regions associated to $AB$, $AC$ and $BC$ are $a\cup b$, $a\cup c$ and $b\cup c$ (or that the inequality (\ref{ineq-ab-acupb}) and the analogue for $a\cup c$ and $b\cup c$ holds), we will have:
\bea\label{q-mmi-bounds}
\int_{AB}u_1\leq S[AB], \qquad \int_{AC}u_2\leq S[AC]\,,\qquad \int_{BC}u_3\leq S[BC]\,,
\eea
provided the following conditions hold
\bea
I[a:b]\leq \int_a s_{ab}+\int_b s_{ba}\,, \quad I[a:c]\leq \int_a s_{ac}+\int_c s_{ca}\,, \quad I[b:c]\leq \int_b s_{bc}+\int_c s_{cb}\,.
\eea
These bounds together imply the monogamy of mutual information for the bulk entropies,
\bea
S[a\cup b]+S[a\cup c]+S[b\cup c]\geq S[a]+S[b]+S[c]+
S[a\cup b\cup c]\,,
\eea
 where in the last line we have change the constraint on the bundle that reproduces $S[d]$ to reproduce instead $S[a\cup b \cup c]$. This is due to fact that in our case the bulk state on $a\cup b\cup c\cup d$ is not pure. In other words, we imposed
  \bea\label{ent-abc}
S[a\cup b\cup c]=\int_{a\cup b\cup c} \(s_{ad}+s_{bd}+s_{cd}\) \quad {\rm instead\,\, of} \quad
S[d]=\int_{d} \(s_{da}+s_{db}+s_{dc}\)\,.
 \eea
The monogamy of bulk entropy is a necessary condition, but from our current derivation it does not seem to be a sufficient one. The geometrization of the bulk entropies, on the other hand, is clearly sufficient but may not be necessary in general as argued in footnote \ref{footnote-16}.

Finally, let us prove boundary monogamy under the above assumptions and constraints.  We start by adding the inequalities (\ref{q-mmi-bounds}) using the relations (\ref{u-flows}), which results in
\bea\label{q-mmi}
S[AB]+S[AC]+S[BC]\geq \int_{A} v_A+\int_B v_B+\int_C v_C-\int_{ABC} v_D\,.
\eea
The first three integrals on the right hand side equal their associated boundary entropies by the quantum max multiflow theorem.
Now, let us work out the last term
\bea
-\int_{ABC} v_D=-\int_{ABC}v^{\rm i}_{D}-\int_{ABC}v^{\rm h}_{D}\,.
\eea
The flux of the homogeneous component on $ABC$ equals minus the flux of the homogeneous component of its complement, which in this case is $D$. Hence
\bea\label{ho-vABC}
-\int_{ABC}v^{\rm h}_{D}=\int_{D}v^{\rm h}_{D}=S_{cl}[D]=S_{cl}[ABC]\,.
\eea
The flux of the inhomogeneous part is
\bea\label{inho-vABC}
-\int_{ABC}v^{\rm i}_{D}=-\int_{a\cup b\cup c} \(s_{da}+s_{db}+s_{dc}\)=\int_{a\cup b\cup c} \(s_{ad}+s_{bd}+s_{cd}\)=S[a\cup b\cup c]\,,
\eea
where we used our modified constraint (\ref{ent-abc}) and the antisymmetry of the entropy bundle. Combining (\ref{ho-vABC}) and  (\ref{inho-vABC}) and plugging them in (\ref{q-mmi}) we arrive at
\bea\label{q-mmi-2}
S[AB]+S[AC]+S[BC]\geq S[A]+S[B]+S[C]+S[ABC]
\eea
which is the monogamy of mutual information.

\section{Perturbative states and Iyer-Wald formalism\label{sec:IyerWald}}

Using tools of convex optimization, we have shown that the quantum corrections to the RT formula can be described in terms of a program involving maximizing the flux of a vector field $v$ with non-trivial sources (\ref{eq:flowFLM}). In this section we will present an application of this prescription to the program of gravitation from entanglement, started in the seminal papers \cite{Lashkari:2013koa,Faulkner:2013ica} and extended in various directions in \cite{Swingle:2014uza,Caceres:2016xjz,Czech:2016tqr,Faulkner:2017tkh,Dong:2017xht,Haehl:2017sot,Lewkowycz:2018sgn,Rosso:2020zkk}. The key idea of this program is to obtain gravitational dynamics from the laws of entanglement entropy in the dual CFT. In \cite{Agon:2020mvu}, these ideas were employed in the context of bit threads. Exploiting the non-uniqueness property of bit threads, \cite{Agon:2020mvu} showed that imposing bulk locality singles out a flow configuration that encodes linearized Einstein's equation for perturbative excited states!

Following \cite{Agon:2020mvu}, here we study in detail the quantum bit thread prescription for the case of perturbative semi-classical states. We will show that the canonical bit thread construction based on the Iyer-Wald formalism, proposed in \cite{Agon:2020mvu}, will naturally incorporate the leading $G_N$ corrections. In particular, we will show that the linearized semi-classical Einstein's equations with sources will be automatically encoded in the perturbed thread configuration.

The Iyer-Wald formalism is naturally expressed in term of differential forms. Therefore, we will start in section \ref{sec:diffform} by giving a very brief overview on how to deal with bit threads in this language and how to translate the different expressions between vector fields and differential forms. Later in section \ref{sec:IW} we will analyze in detail the case of perturbative excited states and explain how the Iyer-Wald formalism comes into play.

\subsection{The language of differential forms\label{sec:diffform}}

The translation of bit threads in terms of differential forms was first presented \cite{Headrick:2017ucz}. Here it was introduced as a way of dealing with the max flow-min cut theorem for null hypersurfaces $\Sigma$, which have degenerate metrics (i.e., with $\det g =0$), rendering the norm bound ill-defined. More recently, this formalism was spelled out in more detail in \cite{Agon:2020mvu}, which introduced the language as a way to get rid of the explicit metric dependence and make the property of bulk locality explicit. Below, we will give a brief summary of the main entries of this dictionary. Along the way, we will emphasize the role of the leading $G_N$ corrections and the differences between the classical and quantum prescriptions. We refer the reader to \cite{Agon:2020mvu} for further details and more detailed explanations about the various statements on differential geometry.

\subsubsection*{Map between vector fields and forms}

In the presence of a metric, a vector field $v$ map to a $(d-1)-$form $\bm w$,
\bea\label{hodgestar}
v^a =g^{a b}(\star \bm w)_{b}\,,\qquad (\star \bm w)_{b}\equiv\frac{1}{(d-1)!} \sqrt{g}\,  w^{a_1\ldots a_{d-1}}\varepsilon_{a_1\ldots a_{d-1} b}\,,
\eea
In the above, $\star \bm w$ represents the Hodge star dual of $\bm w$ and $\varepsilon_{a_1\ldots a_{d}}$ is the totally antisymmetric Levi-Civita symbol, with sign convention $ \varepsilon_{i_1 \ldots i_{d-1}z}=1$. All the indices are raised with the Riemannian metric $g_{ab}$, whose determinant is denoted by $g$. The inverse relation is given by
\bea\label{flow-forms-1}
{\bm w}= \frac{1}{(d-1)!}\epsilon_{a_1 \ldots a_{d-1}b}  \, v^b \,dx^{a_1}\wedge \cdots \wedge dx^{a_{d-1}}\,,
\eea
or, in terms of components,
\bea \label{flow-forms-2}
{w}_{a_1 \ldots a_{d-1}}=  \epsilon_{a_1 \ldots a_{d-1}b} v^b\,.
\eea
Here $ \epsilon_{a_1 \ldots a_{d}}$ are the components of the natural volume form $\bm \epsilon$,
\bea
\bm \epsilon=\frac{1}{d!}\,  \epsilon_{a_1 \ldots a_{d}} dx^{a_1}\wedge \cdots \wedge dx^{a_{d}}\,,
\eea
which are proportional to the components of the Levi-Civita symbol but are normalized such that $\epsilon_{a_1 \ldots a_{d}}=\sqrt{g}\varepsilon_{a_1 \ldots a_{d}}$. Now, taking the exterior derivative of $\bm w$ we find
\bea \label{flows-forms}
d {\bm w}=\(\nabla_a v^a  \) {\bm \epsilon}\,.
\eea
This means that divergenceless vector fields are mapped to closed $(d-1)-$forms, which is relevant to the description of classical bit threads. In the quantum version, we do not need to impose the closedness condition. In fact, the new program (\ref{eq:flowFLM}) requires that
\be\label{quantumclosed}
d {\bm w}=-4G_N s(x){\bm \epsilon}\,,
\ee
so the closedness condition is violated already at linear order in $G_N$.\footnote{This condition could be equivalently expressed in terms of ``generalized calibrations'' \cite{Bakhmatov:2017ihw}. We thank Eoin Colg\'ain for bringing this point to our attention.}

\subsubsection*{Gauss's law and homology condition}

It is also convenient to write down a formula for the restriction of $\bm w$ on a codimension-one surface $\Gamma$. To do so, notice that the
volume $d-$form $\bm \epsilon$ induces a volume $(d-1)-$form $\tilde{\bm \epsilon}$ on $\Gamma$,
\bea\label{inducedeps}
\epsilon_{a_1\ldots a_{d-1}b}=d\,  \tilde{\epsilon}_{[a_1\ldots a_{d-1}}n_{b]}\,,
\eea
where $n$ is the unit normal to $\Gamma$. In terms of this lower-dimensional form, we can thus write
\bea \label{boundaryw}
\bm w|_{\,\!_\Gamma}=(n_a v^a) \tilde{\bm \epsilon}\,.
\eea
With the above ingredients we can now consider Gauss's theorem applied to the bulk region $\Sigma_A$ with $\partial \Sigma_A= A\cup \gamma_A$ ($\gamma_A\sim A$):
\bea
\int_{\Sigma_{A}}\nabla_a v^a \bm{\epsilon}=-4G_N\int_{\Sigma_{A}}s(x) \bm{\epsilon}=\int_{\partial \Sigma_{A}}\!\! \(n_a v^a\) \tilde{\bm \epsilon}=\int_{\gamma_A} \(n_a v^a\) \tilde{\bm \epsilon}-\int_{A} \(n_a v^a\) \tilde{\bm \epsilon}\,,
\eea
which leads to the homology condition
\bea\label{homo-cond}
\int_{A}\(n_a v^a\) \tilde{\bm \epsilon}=\int_{\gamma_A}\!\!\(n_a v^a\) \tilde{\bm \epsilon}+4G_N S_{\text{bulk}}(\Sigma_{A})\,.
\eea
We can arrive to an analogous statement in the language of forms. Using Stoke's theorem:
\bea
\int_{\Sigma_{A}} d {\bm w}=-4G_N\int_{\Sigma_{A}}s(x) \bm{\epsilon}=\int_{\partial \Sigma_{A}} \!\!{\bm w}=\int_{\gamma_A} {\bm w}-\int_{A} \!\!{\bm w}\,,
\eea
which leads to
\bea\label{homcondiff}
\int_{A} {\bm w}=\int_{\gamma_A}\!\!{\bm w}+4G_N S_{\text{bulk}}(\Sigma_{A})\,.
\eea
In summary, since the closedness condition is violated by quantum corrections (\ref{quantumclosed}), Gauss's law picks up a volume term
that captures the sources of entanglement in the bulk. The integral over these sources yield $S_{\text{bulk}}(\Sigma_A)$, hence, the surface integrals
over $A$ and $\gamma_A$ end up differing precisely by this factor, which is of order $\mathcal{O}(G_N)$.

\subsubsection*{Max flow-min cut theorem and quantum corrections}

Finally, we can translate the max flow-min cut theorem in terms of differential forms and rewrite the quantum corrected dual program in this language. First, notice that the form $\bm w$ evaluated on any codimension-one hypersurface $\gamma_A$ is bounded by its natural volume form $ \tilde{\bm \epsilon}$,
\bea\label{eq:boundw}
\int_{\gamma_A} \bm w = \int_{\gamma_A} \(n_a v^a\) \tilde{\bm \epsilon}\leq  \int_{\gamma_A} \tilde{\bm \epsilon}\,.
\eea
This inequality is a consequence of the norm bound, $|v|\leq1$. In terms of differential forms the latter can be written as
\be\label{w-norm1}
\langle\bm w,\bm w\rangle_{g}\leq1\,,
\ee
where $\langle \cdot,\!\cdot \rangle_g$ denotes the inner product in the presence of a metric
\bea\label{innerp}
\langle \bm x , \bm y\rangle_g=\frac{1}{(d-1)!}g^{a_1 b_1}\cdots g^{a_{d-1} b_{d-1}} x_{a_1 \ldots a_{d-1}}y_{b_1 \ldots b_{d-1}}\,.
\eea
It also implies that an optimal form $\bm w^*$, which saturates the inequality in (\ref{eq:boundw}), should be equivalent to volume form $\tilde{\bm \epsilon}$ at the surface $\gamma_A$, i.e.,
\bea\label{formmA}
{\bm w^*}|_{\gamma_A}=\tilde{\bm \epsilon}|_{\gamma_A}\,.
\eea
Now, we add $4G_N S_{\text{bulk}}(\Sigma_{A})$ to both sides of the inequality (\ref{eq:boundw}) and use the homology condition (\ref{homcondiff}) to write
\be
\int_{A} \bm w\leq  \int_{\gamma_A} \tilde{\bm \epsilon}+4G_N S_{\text{bulk}}(\Sigma_{A})\,,
\ee
so that the LHS no longer makes reference to the surface $\gamma_A$. The max flow-min cut theorem is obtained by maximizing the LHS of (\ref{eq:boundw}) and minimizing its RHS for arbitrary homologous surfaces $\gamma_A$.
This leads to
\bea\label{eq:MFMC}
\underset{\bm w \in \bm W}{\rm max} \int_{A} \bm w =\underset{\gamma_A\sim A}{\rm min } \left(\int_{\gamma_A}\!\! \tilde{\bm \epsilon}+4G_N S_{\text{bulk}}(\Sigma_{A})\right)\,,
\eea
where $\bm W$ is the set of forms obeying (\ref{quantumclosed}) and (\ref{w-norm1}). The RHS of this equation looks like the QES prescription. However, we need to  point out that the condition (\ref{quantumclosed}) implies that a local $s(x)$ is known, and is associated with the $\gamma_A$ that arises from the minimization in the RHS. To avoid circular reasoning we must therefore relax the second step and drop the $S_{\text{bulk}}$ term from the minimization, i.e.,
\be\label{preFLM}
\underset{\bm w \in \bm W}{\rm max} \int_{A} \bm w =\underset{\gamma_A}{\rm min } \int_{\gamma_A}\!\! \tilde{\bm \epsilon}+4G_N S_{\text{bulk}}(\Sigma_{A})\,,
\ee
which can be justified if $G_N$ is perturbatively small.
The RHS of (\ref{preFLM}) now looks like the FLM formula which yields the first correction in $G_N$ over the RT formula. Alternatively, we can keep the full minimization in (\ref{eq:MFMC}) but bearing in mind that the results are only valid at leading order in $G_N$. This can be done consistently, as was explained in the previous section. In fact, there are good reasons do do so, since the FLM and QES prescriptions can give different results at this order in situations close to a phase transition and only the latter that gives the correct result there. Putting everything together, then, we arrive at
\bea\label{QBTconditionsF}
S_A=\frac{1}{4G_N}\,\underset{\bm w \in \bm W}{\rm max}\,\, \int_{A} \bm w\,,\qquad\bm W=\left\{\bm w \,|\, d\bm w =-4G_N s(x)\bm \epsilon\,, \langle\bm w,\bm w\rangle_{g}\leq1 \right\}\,.
\eea
This is the differential form version of the max-flux formula (\ref{eq:flowFLM}).

\subsubsection{Perturbative quantum states\label{sec:pert}}

Let us now consider the class of perturbative excited states $|\psi\rangle_{\text{CFT}}$ with semi-classical gravity duals. This kind of states can be described in the bulk by specifying a pair $\{M,|\psi\rangle_{\text{bulk}}\}$ consisting of a classical manifold $M$ and a bulk quantum state living on this manifold $|\psi\rangle_{\text{bulk}}$. The quantum fields in the bulk will generally be in an excited state as well, so they will induce a non trivial stress tensor $\langle\psi| T_{\mu\nu}^{\text{bulk}}|\psi\rangle\neq0$, which enters as a source of the semi-classical Einstein's equations
\be\label{EE-semicl}
R_{\mu\nu}-\frac{1}{2}R g_{\mu\nu}+\Lambda g_{\mu\nu}=8\pi G_N\langle\psi| T_{\mu\nu}^{\text{bulk}}|\psi\rangle\,.
\ee
For perturbative excited states, however, the energy of the bulk fields is not very large and we can treat the backreaction perturbatively,
\be\label{pert:metric}
g_{\mu\nu}=g_{\mu\nu}^{(0)}+\delta g_{\mu\nu}\,,
\ee
where $g_{\mu\nu}^{(0)}$ is the metric of pure AdS and $\delta g_{\mu\nu}$ is the first order correction. Even though (\ref{pert:metric}) can be regarded as a linear perturbation of the metric, we emphasize that the perturbative states we are considering here are very different in nature to the states that were studied in \cite{Agon:2020mvu}. The main difference is that the states considered in that paper were dual to classical vacuum perturbations in the bulk, implying that $\langle T_{\mu\nu}^{\text{bulk}}\rangle=0$ and $\delta g_{\mu\nu}\sim\mathcal{O}(G_N^0)$. In our case, however, $\delta g_{\mu\nu}$ is understood to arise as a result of the backreaction due to the quantum fields, so we expect $\delta g_{\mu\nu}\sim\mathcal{O}(G_N)$.\footnote{For both kind of states we still require that $\delta g_{\mu\nu}\ll g_{\mu\nu}^{(0)}$, but the difference is the nature of the ``small parameters'' in which the perturbation expansions are carried out.} Another point to emphasize is that, given a stress tensor in the bulk $\langle T_{\mu\nu}^{\text{bulk}}\rangle$, $\delta g_{\mu\nu}$ is not uniquely determined, as we can always add a \emph{homogeneous} solution of the linearized equations of motion $\delta g^{\text{(H)}}_{\mu\nu}$ to a particular \emph{inhomogeneous} solution $\delta g^{\text{(I)}}_{\mu\nu}$, i.e., $\delta g_{\mu\nu}\to\delta g^{\text{(I)}}_{\mu\nu}+\delta g^{\text{(H)}}_{\mu\nu}$ and it will still satisfy the same equations. However, if we impose proper boundary conditions on $\partial M$, then the combination $\delta g^{\text{(I)}}_{\mu\nu}+\delta g^{\text{(H)}}_{\mu\nu}$ indeed turns out to be unique. This will be important for the discussion below, specifically, for the claims addressing the existence and uniqueness of the proposed canonical thread configurations.

Now, given the above information, i.e., a manifold $M$ with metric $g_{\mu\nu}^{(0)}+\delta g_{\mu\nu}$ and a bulk state $|\psi\rangle_{\text{bulk}}$, we would now like to understand the implications for the max flux problem. We will denote a solution to such problem as $\bm w$, and split it as
\be
\bm w=\bm w^{(0)}+\delta \bm w\,,
\ee
where $\bm w^{(0)}$ is the solution of the max flux problem in the pure AdS case (or $G_N\to0$ limit) and $\delta \bm w$ is the perturbation, which we can expect to be of order $ \mathcal{O}(G_N)$. Now, the condition (\ref{quantumclosed}) implies
\bea\label{ddelta}
d( \bm w ^{(0)}+\delta \bm w )=-4G_N s(x){\bm \epsilon}\,,
\eea
which can be split as
\be\label{ddeltasplit}
 d\bm w ^{(0)}=0\,,\qquad d (\delta \bm w )=-4G_N s(x){\bm \epsilon}\,.
\ee
Similar to the metric, $\delta \bm w$ can only be specified once we impose a boundary condition on $\partial M$. The reason is that without such a boundary condition
one can always add a closed form and write a more general solution as $\delta \bm w=\delta \bm w^{\text{(I)}}+\delta \bm w^{\text{(H)}}$, with $d(\delta \bm w^{\text{(I)}})=-4G_N s(x){\bm \epsilon}$ and $d(\delta \bm w^{\text{(H)}})=0$. 
On the other hand, the minimal surface $\gamma_A$ does not change at first order in the perturbation. Since this is
a bottle-neck for the flow, $\bm w$ is fixed at its location according to (\ref{formmA}), i.e.,
\bea\label{deltaboundary}
(\bm w ^{(0)}+ \delta \bm w)|_{\gamma_A}=\tilde{\bm \epsilon }|_{\gamma_A}=(\tilde{\bm \epsilon}^{(0)}+\delta \tilde{\bm \epsilon })|_{\gamma_A}\,,
\eea
which, similarly, can be split as
\be\label{bcformpert}
\bm w ^{(0)}|_{\gamma_A}=\tilde{\bm \epsilon }^{(0)}|_{\gamma_A}\,,\qquad \delta \bm w |_{\gamma_A}=\delta \tilde{\bm \epsilon }|_{\gamma_A}\,,
\ee
Finally, the norm bound (\ref{w-norm1}) translates into
\bea\label{P-norm-bound}
\langle \bm w ^{(0)}, \bm w^{(0)}\rangle_g+\[2\langle \bm w ^{(0)}, \delta \bm w\rangle_g +\langle \bm w ^{(0)}, \bm w^{(0)}\rangle_{\delta g}\] \leq 1\,,
\eea
where $\langle \cdot,\! \cdot\rangle_{g}$ is the usual inner product defined in (\ref{innerp}) and $\langle \cdot,\! \cdot\rangle_{\delta g}$ denotes
\bea
\langle \bm x , \bm y\rangle_{\delta g}=\frac{1}{(d-1)!}\delta(g^{a_1 b_1}\ldots g^{a_{d-1} b_{d-1}}) x_{a_1 \ldots a_{d-1}}y_{b_1 \ldots b_{d-1}}\,.
\eea
The leading order term in (\ref{P-norm-bound}) satisfies
\be\label{bound0th}
\langle \bm w ^{(0)}, \bm w^{(0)}\rangle_g\leq 1\,.
\ee
This bound it is already saturated (at least) at the location of the bulk bottle-neck $\gamma_A$. The second term is suppressed by a factor of order $\mathcal{O}(G_N)$ so it is clear that (\ref{P-norm-bound}) it is only in danger whenever (\ref{bound0th}) is saturated or very close to saturation. In general, the norm bound will typically depend on $\bm w^{(0)}$ so a priori it seems unlikely that a generic $\delta \bm w$ obeying  (\ref{ddeltasplit}) and (\ref{bcformpert}) could satisfy (\ref{P-norm-bound}) independent of $\bm w^{(0)}$. Fortunately, (\ref{bcformpert}) already implies that the bracket in (\ref{P-norm-bound}) vanishes at $\gamma_A$ so it is possible to satisfy (\ref{P-norm-bound}) provided we pick a $\bm w^{(0)}$ that decays rapidly away from the minimal surface \cite{Agon:2020mvu}. In the remaining part of this section we will show that, in fact, the Iyer-Wald formalism provides a concrete realization for $\delta \bm w$ satisfying all the above conditions.

\subsection{Canonical bit threads from Iyer-Wald\label{sec:IW}}

One of the most exciting developments in the context of AdS/CFT is the program of \emph{gravitation from entanglement}, initiated in the seminal papers \cite{Lashkari:2013koa,Faulkner:2013ica}. This program aims to connect the dynamical equations of motion in the bulk, i.e., the Einstein's equations, with the laws that govern the dynamics of entanglement entropy in the dual CFT. This map was initially proved at the linearized level, by directly comparing the equations in both sides and making use of the known entries of the AdS/CFT dictionary \cite{Lashkari:2013koa}. A more elegant proof of this map was worked out in \cite{Faulkner:2013ica} using the Iyer-Wald formalism, which is well-known for relativists and is widely used in the context of black hole thermodynamics. More recently, this program was developed and extended in various directions in \cite{Swingle:2014uza,Caceres:2016xjz,Czech:2016tqr,Faulkner:2017tkh,Dong:2017xht,Haehl:2017sot,Lewkowycz:2018sgn,Rosso:2020zkk} and was translated into the language of bit threads in \cite{Agon:2020mvu}, in the regime where quantum corrections are suppressed. The outcome of this work was a concrete proposal for a bit thread configuration that satisfies all the requirements for a divegenceless flow \emph{and} make use of the property of bulk locality, encoding gravitational Einstein's equations linearized over AdS. Among other things, the \cite{Agon:2020mvu} showed that such a \emph{canonical} choice for the thread configuration could be used efficiently for the problem of bulk reconstruction, giving rise to explicit formulas for the metric as the inverse of certain operator that can be specified entirely from CFT data.

In this section we will show that the framework developed in \cite{Agon:2020mvu} can, in fact, accommodate for the leading $G_N$ corrections when considering semi-classical states in the bulk. We will take inspiration from \cite{Swingle:2014uza} which already extended the results of \cite{Faulkner:2013ica} to include the leading quantum corrections, hence, showing that the semi-classical Einstein's equations universally coupled to matter emerge from the dynamics of entanglement entropy in the boundary CFT. We will show that, in the context of quantum bit threads, this statement translates into a specific proposal for a bit thread configuration that encodes the leading quantum corrections while making use of the property of bulk locality. 

\subsubsection{Classical excited states and vacuum Einstein's equations\label{sec:IW1}}

Let us start by reviewing the main results of \cite{Agon:2020mvu}, regarding. In general QFTs (holographic or not),
for small perturbations over a reference state, $\rho=\rho^{(0)}+ \delta\rho$, entanglement entropy satisfies the so-called first law of entanglement,
\be\label{eq:1stlaw}
\delta S_A=\delta\langle H_A\rangle\,,
\ee
where $\langle \cdot \rangle$ represents the expectation value of the operator in the respective quantum state and $H_A$ is known as the entanglement or modular Hamiltonian. The latter operator is formally related to the reduced density matrix $\rho_A=\text{tr}_{\bar{A}}[\rho]$ through
\be\label{defmodularH}
\rho_A=\frac{e^{-H_A}}{\text{tr}[e^{-H_A}]}\,.
\ee
However, there are very few instances in which (\ref{defmodularH}) can be explicitly solved to obtain $H_A$. A simple case where such inversion is possible is when $A$ is taken to be half-space, say $x_1>0$, and $\rho$ corresponds to the vacuum state of the QFT. In this case $H_A$ corresponds to the generator of time-translations associated with a family of Rindler observers that can perform measurements in the region $x_1>0$ \cite{Bisognano:1975ih,Unruh:1976db}
\begin{equation}\label{modular1}
H_A=2\pi\int_A x_1 \, T_{00}(t,\vec{x}) \, d^{d-1}x\,.
\end{equation}
For generic CFTs, this setup can be conformally mapped to the case where
$A$ is a ball of radius $R$, centered at an arbitrary point $\vec{x}=\vec{x}_c$, in which case \cite{Hislop:1981uh,Casini:2011kv}
\begin{equation}\label{modular2}
  H_A=2\pi\int_A  \frac{R^2-(\vec{x}-\vec{x}_c)^2}{2R} T_{00}(t,\vec{x})\,d^{d-1}x\,.
\end{equation}
Again, in this case $H_A$ can be interpreted as the generator of time-translations for a particular class of Rindler observers that have only access to the interior of the ball. This means that, at least for these special cases, (\ref{eq:1stlaw}) can be recast as a \emph{thermodynamic} first law,
\be\label{1stlawtherm}
\delta S_A=\delta E_A\,,
\ee
where $\delta E_A$ is some energy associated with the subsystem $A$. The key distinction here is that (\ref{eq:1stlaw}) applies more generally, i.e., it does not require that we are varying to a nearby equilibrium state. It is an exact quantum relation rather than a thermodynamic one.

The main insight of \cite{Lashkari:2013koa,Faulkner:2013ica} was to show that the first law (\ref{eq:1stlaw}), applied to a family of ball-shaped regions in the boundary, is in one-to-one correspondence with the \emph{homogeneous} Einstein's equations in the bulk, linearized over pure AdS. The key ingredient of their proof (in the language of \cite{Faulkner:2013ica}) was to show that there is a $(d-1)-$form $\tilde {\bm \chi}$ which satisfies
\be\label{propschi}
\int_A \tilde{\bm \chi}=\delta \langle H_A\rangle=\delta E_A^{\text{grav}}\,,\qquad \int_{\gamma_A}\tilde{\bm \chi}=\frac{1}{4G_N}\int_{\gamma_A}\delta\tilde{\bm \epsilon}=\delta S_A^{\text{grav}}\,,
\ee
and
\be\label{}
d\tilde{\bm \chi}=-2\xi^{\mu}\delta E^{g}_{\mu\nu}{\bm \epsilon}^\nu\,,
\ee
where $\delta E_A^{\text{grav}}$ and $\delta S_A^{\text{grav}}$ are the gravitational versions of the quantities $\delta E_A$ and $\delta S_A$ in (\ref{1stlawtherm}) that arise upon applying the holographic dictionary, $\delta E^{g}_{\mu\nu}$ are the \emph{linearized} version of Einstein's equations without any bulk matter, i.e.,
\be\label{linearizedEab}
E^{g}_{\mu\nu}=\frac{1}{\sqrt{-g}}\frac{\delta \mathcal{S}_g}{\delta g^{\mu\nu}}\,,\qquad \mathcal{S}_g\equiv\frac{1}{16\pi G_N}\int d^{d+1}x\sqrt{-g}(R-2\Lambda)\,,
\ee
${\bm \epsilon}^\nu$ is the volume form on $\Sigma$, and $\xi^{\mu}$ is a time-like conformal Killing vector given by (in Poincar\'e coordinates)
\bea
\xi=-\frac{2\pi}{R}\(t-t_0\)[z\partial_z+(x^i-x^i_0)\partial_i]+\frac{\pi}{R}[R^2-z^2-(t-t_0)^2-(\vec{x}-\vec{x}_0)^2]\partial_t\,.
\eea
When considering on-shell perturbations, one has that $\delta E^{g}_{\mu\nu}=0$ which implies that $\tilde{\bm \chi}$ is closed. Stoke's theorem then implies that the first law of entanglement is satisfied, $\delta S_A^{\text{grav}}=\delta E_A^{\text{grav}}$. Conversely, starting from the first law of entanglement and using Stoke's theorem one finds that $d\tilde{\bm \chi}$ integrates to zero in the bulk region bounded by $A\cup\gamma_A$. Considering all possible balls $A$ in all Lorentz frames then leads to $\delta E^{g}_{\mu\nu}=0$ at each bulk point \cite{Faulkner:2013ica}, i.e., the linearized Einstein's equations without matter.

The above observations led to the proposal that $\tilde{\bm{\chi}}$ could be a good object to consider for the construction of the perturbed $(d-1)-$form $\delta \bm w$. More specifically, \cite{Agon:2020mvu} argued that we could identify
\bea \label{w-chi}
\delta \bm w=4G_N\, \tilde{\bm{\chi}}\,,
\eea
when considering perturbations of AdS \emph{without} matter. This proposal for $\delta \bm w$ immediately satisfy two of the required properties: i) the fact that $d(\delta \bm w)=0$, at least for on-shell vacuum perturbations, and ii) the fact that it satisfies the boundary condition (\ref{deltaboundary}), implied by the second equation in (\ref{propschi}). The final condition that still needed to be checked was the norm bound (\ref{P-norm-bound}). However, \cite{Agon:2020mvu} showed that there always exists a generic $(d-1)-$form $\bm w^{(0)}$ that implies the latter. The specific $\bm w^{(0)}$ considered in that paper was an example of a \emph{geodesic flow}, constructed originally in \cite{Agon:2018lwq}. Indeed, such solutions have the special property of decaying fast enough away from the minimal surface $\gamma_A$ so that (\ref{P-norm-bound}) is guaranteed for all points in $M$. Notice that, the fact that $\delta\bm w$ is closed only when we are dealing with on-shell vacuum perturbations implies that $d(\delta \bm w)=0$ secretly encodes the Einstein's equations of motion, which is a nice property of this particular construction. In particular, \cite{Agon:2020mvu} showed that this feature could be exploited to address the question of metric reconstruction for excited states, eluding various limitations of alternative methods.

As a final remark, we emphasize that (\ref{w-chi}) is strictly valid for \emph{classical} perturbative excited states with no sources in the bulk. In this case we have assumed that $\delta g_{\mu\nu}\ll g_{\mu\nu}^{(0)}$ but $\delta g_{\mu\nu}$ is taken to be of order $\mathcal{O}(G_N^0)$. In the next section we will analyze what happens for semi-classical perturbative excited states. In that case we will also have that $\delta g_{\mu\nu}\ll g_{\mu\nu}^{(0)}$ but now with $\delta g_{\mu\nu}\sim\mathcal{O}(G_N)$. Hence, besides the area term for $\delta S_A^{(\text{grav})}$ appearing in (\ref{propschi}), we will also need to include an additional term that enters at the same order, which is due to the entanglement entropy of bulk fields. Before proceeding with this analysis, let us first offer a couple of comments that will help to clarify the physical meaning of (\ref{w-chi}) and highlight the role of the Iyer-Wald formalism. This will in turn shed light on the generalization of this result for the semi-classical states that we will consider next.

First, we point out that Killing vector $\xi$ has a bifurcate horizon precisely at $\gamma_A$, which in Poincar\'e coordinates is given by the collection of points on a hemisphere of radius $R$,
\be
\gamma_A=\{(z,\vec{x})\,|\, R^2=z^2+|\vec{x}|^2 \}\,.
\ee
The flow associated to this Killing vector corresponds to a natural class of Rindler observers in the CFT associated with the region $A$ and a corresponding class of bulk observers for which the horizon $\gamma_A$ represents the boundary of their knowledge of the bulk state. In fact, a specific conformal transformation (known as the CHM map \cite{Casini:2011kv}) maps the interior of the entanglement wedge associated with $A$ to the exterior of an hyperbolic black hole in AdS. In this conformal frame the killing vector $\xi$ coincides with the generator of time translations $\partial_\tau$, so the first law of entanglement (\ref{1stlawtherm}) translates literally into the first law of \emph{thermodynamics} for this black hole. Following Iyer and Wald \cite{Iyer:1994ys,Iyer:1995kg,Wald:2005nz}, then, one can expect to understand such a first law as a result of applying Noether's theorem for the Killing symmetry generated by $\xi$. Indeed, by doing so one ends up defining the $(d-1)-$form
\bea\label{chi}
{ \bm \chi}=-\frac{1}{16 \pi G_N} \left[ \delta (\nabla^\mu \xi^\nu {\bm \epsilon}_{\mu\nu} )+\xi^\nu {\bm \epsilon}_{\mu\nu}(\nabla_\sigma \delta g^{\mu\sigma}+\nabla^\mu \delta g^\sigma_{\,\, \sigma})\right]\,,
\eea
where $\delta g_{\mu\nu}$ denotes the metric perturbation, as in (\ref{pert:metric}), and ${\bm \epsilon}_{\mu\nu}$ is the volume $(d-1)-$form
\bea
{\bm \epsilon}_{\mu\nu}=\frac{1}{(d-1)!}\epsilon_{\mu\nu\sigma_3 \cdots \sigma_{d+1}}dx^{\sigma_3} \wedge\cdots \wedge dx^{\sigma_{d+1}}\,,
\eea
with $\epsilon_{z t i_1 \cdots i_{d-1}}=\sqrt{-g}$. To get a handle on the above$(d-1)-$form, it is useful to completely fix the gauge for the metric perturbations and specialize ${\bm \chi}$ to a Cauchy slice
$\Sigma$ containing the bifurcate horizon $\gamma_A$. For instance, working in the Fefferman-Graham gauge, where $h_{\mu\nu}=z^{d-2}H_{\mu\nu}$ (for $\mu,\,\nu\in\{0,i\}$) and $h_{zz}=h_{z\mu}=0$ (in Poincar\'e coordinates),
and taking $\Sigma$ to be the $t=t_0$ slice, one obtains the $(d-1)-$form
\bea\label{chiSigma}
{ \bm \chi}|_{\,\!_\Sigma}\equiv\tilde{\bm {\chi}} &=&\frac{z^d}{16 \pi G_N} \Bigg\{{\bm \epsilon}^t_{\,\,z}\left[ \(\frac{2\pi z}{R}+\frac{d}{z} \xi^t+\xi^t\partial^z \) H^i_{\,\,i} \right] +\nonumber \\ \label{chi-Sigma}
 && + {\bm \epsilon}^t_{\,\, i} \left[ \(\frac{2\pi (x^i-x^i_0)}{R}+\xi^t\partial^i \) H^j_{\,\,j} -\(\frac{2\pi (x^j-x^j_0)}{R}+\xi^t\partial^j\) H^i_{\,\,j} \right]  \Bigg\}\,.
\eea
With (\ref{chiSigma}) at hand, is then easy to check that both equations in (\ref{propschi}) are indeed satisfied. Notice that in the Iyer-Wald formalism $S_A^{\text{grav}}$ is interpreted as the Noether charge associated with the Killing symmetry generated by $\xi$.

\subsubsection{Quantum excited states and semi-classical Einstein's equations\label{sec:IW2}}

Let us now discuss how to deal with semi-classical bulk states. Our discussion will follow closely \cite{Swingle:2014uza} which used the results of \cite{Faulkner:2013ica} to investigate this kind of states, and showed that the semi-classical Einstein's equations universally coupled to matter (\ref{EE-semicl}) emerge from the same CFT equation (\ref{1stlawtherm}). Our task is then to review their argument and rewrite it in the language of quantum bit threads.

As in \cite{Faulkner:2013ica}, it will be important to make sure to include all subleading terms that enter at the first subleading order in the perturbation for the various entries of the holographic dictionary. We will start by discussing the expected corrections for $\delta S_A^{\text{grav}}$ and $\delta E_A^{\text{grav}}$, respectively, and then move to the argument for the semi-classical Einstein's equations.

\subsubsection*{i) Dictionary for $\delta S_A^{\text{grav}}$}

There are two contributions to $\delta S_A^{\text{grav}}$, one due to the change in the area and another one due to the entanglement of bulk fields, both of which are of order $\mathcal{O}(G_N^0)$. To understand this, let us recall that the class of CFT states that we are considering are described in the bulk by a classical manifold $M$ and a bulk quantum state living on this manifold $|\psi\rangle_{\text{bulk}}$. We are specifically interested on the case of perturbative excited states. Working directly within the bulk Hilbert space, this means that we want to focus on states of the form
\be\label{excitedstates}
|\psi\rangle_{\text{bulk}}=|0\rangle+\lambda\sum_{k_i}(c_{k_1} a^\dag_{k_1}+c_{k_1k_2} a^\dag_{k_1}a^\dag_{k_2}+\cdots)|0\rangle\,,
\ee
where $a^\dag_{k_i}$ are creation operators and $\lambda$ is the small parameter defining the infinitesimal variation. In this case, it can be shown that the bulk stress energy tensor receives contributions that are of order $\mathcal{O}(\lambda)$. In particular, the quadratic part will include terms of the form $a^\dagger_{k_1}a^\dagger_{k_2}$ and $a_{k_1}a_{k_2}$, in addition to $a_{k_1}a^\dagger_{k_2}$ terms, so its expectation value will be in general nonvanishing,
\be
\langle \psi|T_{\mu\nu}^{\text{bulk}}|\psi\rangle\neq0\,.
\ee
This stress energy tensor will enter as a source of the semi-classical Einstein's equations (\ref{EE-semicl}) and thus, the backreaction will induce a change in the metric according to (\ref{pert:metric}), i.e., $g_{\mu\nu}=g_{\mu\nu}^{(0)}+\delta g_{\mu\nu}$, with $\delta g_{\mu\nu}\sim\mathcal{O}(G_N)$. The surface $\gamma_A$ will not change at linear order in the perturbation, however, the volume form will change according to $\bm \epsilon=\bm \epsilon^{(0)}+\delta\bm \epsilon$, with $\delta\bm \epsilon\sim\mathcal{O}(G_N)$, and similarly for its projection on $\gamma_A$, $\bm \epsilon|_{\gamma_A}=\tilde{\bm \epsilon}=\tilde{\bm \epsilon}^{(0)}+\delta\tilde{\bm \epsilon}$. This will induce a correction to the area
\be
\text{Area}(\gamma_A)=\int_{\gamma_A}\tilde{\bm \epsilon}=\int_{\gamma_A}\tilde{\bm \epsilon}^{(0)}+\int_{\gamma_A}\delta\tilde{\bm \epsilon}\,,
\ee
so that
\be\label{deltaArea}
\frac{\delta \text{Area}(\gamma_A)}{4 G_N}=\frac{1}{4G_N}\int_{\gamma_A}\delta\tilde{\bm \epsilon}\sim\mathcal{O}(G_N^0)\,.
\ee

We note that for the classical perturbative states considered in the previous section the correction in the area in (\ref{deltaArea}) is instead of order $\mathcal{O}(1/G_N)$. Hence, the bulk entanglement entropy, which is of order $\mathcal{O}(G_N^0)$, is highly suppressed and can be sensibly neglected. However, for semi-classical states both terms enter at the \emph{same order} in a $G_N$ expansion, i.e., $\mathcal{O}(G_N^0)$, and one should correct the dictionary so that
\be\label{changeSAquantum}
\delta S_A^{\text{grav}}=\frac{\delta \text{Area}(\gamma_A)}{4 G_N}+\delta S_{\text{bulk}}(\Sigma_A)=\frac{1}{4G_N}\int_{\gamma_A}\delta\tilde{\bm \epsilon}+\delta S_{\text{bulk}}(\Sigma_A)\,.
\ee
In addition, we will exploit the fact that, for bulk regions with a local modular Hamiltonian, the bulk entropy can be computed as
\be\label{bulkMH}
\delta S_{\text{bulk}}=\langle \hat{K}_{\text{bulk}}\rangle=\int_{\Sigma_A}\!\! \xi^\mu \langle T^{\text{bulk}}_{\mu\nu}(x)\rangle {\bm \epsilon}^\nu\,.
\ee
Together with (\ref{changeSAquantum}), this implies that
\be
\delta S_A^{\text{grav}}=\frac{1}{4G_N}\int_{\gamma_A}\delta\tilde{\bm \epsilon}+\int_{\Sigma_A}\!\! \xi^\mu \langle T^{\text{bulk}}_{\mu\nu}(x)\rangle {\bm \epsilon}^\nu\,,
\ee
or, in terms of the form $\tilde{\bm \chi}$ defined via the Iyer-Wald formalism,
\be\label{deltaSIW}
\delta S_A^{\text{grav}}=\int_{\gamma_A}\tilde{\bm \chi}+\int_{\Sigma_A}\!\! \xi^\mu \langle T^{\text{bulk}}_{\mu\nu}(x)\rangle {\bm \epsilon}^\nu\,.
\ee

\subsubsection*{ii) Dictionary for $\delta E_A^{\text{grav}}$}

It can be shown that the expression for $\langle T_{\alpha\beta}^{\text{CFT}}\rangle$, and hence $\delta E_A^{\text{grav}}$, remain unaffected as long as we restrict to bulk perturbations that decay quick enough near the boundary, which we will assume for the sake of simplicity. To see this, we need to do a precise near-boundary analysis for the theory we are interested in.
As a concrete example, let us consider a scalar field theory minimally coupled to Einstein gravity,
\be
S=\frac{1}{16\pi G_N}\int d^{d+1}x\sqrt{-g}\left[R+\frac{d(d-1)}{L^2}\right]-\int d^{d+1}x\sqrt{-g}\left[\frac{1}{2}(\partial\phi)^2+\frac{m^2}{2}\phi^2\right]\,.
\ee
The mass of the scalar field $m$ can be related to the conformal dimension of the dual operator $\mathcal{O}_\Delta$. In the standard quantization, this relation is given by
\be
\Delta=\frac{d+\sqrt{d^2+4m^2L^2}}{2}\,.
\ee
In the quantization of the scalar field one needs to impose boundary conditions for the wave functions, which can be done by specifying the coefficient of the non-normalizable mode
\be
\phi|_{\partial M}\to \phi_{d-\Delta}z^{d-\Delta}\,.
\ee
Killing off the source term, i.e., setting $\phi_{d-\Delta}=0$, then leads to wave functions $\phi_k$ that decay as slowly as possible near the boundary, i.e., $\phi_k\sim\mathcal{O}(z^\Delta)$. The subindex $k$ here denotes collectively a set of quantum numbers, which will depend on the number of dimensions $d$. With these wave functions at hand, one can write off a mode expression for $\phi$ in terms of creation and annihilation operators,
\be
\phi=\sum_{k}\left(e^{-i\omega_k t}\phi_k a_k+e^{i\omega_k t}\phi_k^{*} a_k^\dagger\right)\,,
\ee
and one can then construct the bulk Hilbert space, which will typically contain states $|\psi_{\text{bulk}}\rangle$ of the form (\ref{excitedstates}). This leads to bulk stress tensors that decay near the boundary as
\be
\langle \psi|T_{\mu\nu}^{\text{bulk}}|\psi\rangle \sim \mathcal{O}(z^{2\Delta-d})\,,
\ee
and, via the semi-classical Einstein's equation (\ref{EE-semicl}), to a backreacted metric of the form
\be
ds^2=\frac{L^2}{z^2}\left[dz^2+(\eta_{\alpha\beta}+z^{d}\tau_{\alpha\beta}^{(1)}(x^{\sigma})+z^{2\Delta}\tau_{\alpha\beta}^{(2)}(x^{\sigma})+\cdots)dx^{\alpha}dx^{\beta}\right]\,,
\ee
where we have indicated the explicit dependence with $z$. For $\Delta>\frac{d}{2}$, the term proportional to $\tau_{\alpha\beta}^{(1)}$ gives the leading normalizable behavior, and the GKPW recipe \cite{Gubser:1998bc,Witten:1998qj} leads to the standard result (after appropriate holographic renormalization \cite{Balasubramanian:1999re,deHaro:2000vlm})
\be\label{CFTTmn}
\langle T_{\alpha\beta}^{\text{CFT}}\rangle=\frac{d L^{d-1}}{16\pi G_N}\tau_{\alpha\beta}^{(1)}(t,\vec{x})\,,
\ee
so
\be\label{deltaEcorr}
\delta E_A^{\text{grav}}=\frac{d L^{d-1}}{8 G_N}\int_A  \frac{R^2-(\vec{x}-\vec{x}_c)^2}{2R} \tau_{\alpha\beta}^{(1)}(t,\vec{x})\,d^{d-1}x=\delta\langle H_A\rangle\,.
\ee
The term proportional to $\tau_{\alpha\beta}^{(2)}$ dominates if the operator is sufficiently relevant, i.e. for $\frac{d}{2}-1<\Delta<\frac{d}{2}$. In this case the recipe for $\langle T_{\alpha\beta}^{\text{CFT}}\rangle$ changes so will not consider these cases here. Nevertheless, an analogous formula to (\ref{CFTTmn}) could be obtained by properly doing holographic renormalization, either in the standard or alternative quantization (see e.g. \cite{Casini:2016rwj}). It would be interesting to see the effects of these modified prescriptions for $\delta E_A^{\text{grav}}$ in our context, however, we will leave this exploration to be considered elsewhere. Finally, since the expression (\ref{deltaEcorr}) does not get any correction, in terms of the form $\tilde{\bm \chi}$ we can still write
\be\label{deltaEIW}
\delta E_A^{\text{grav}}=\int_A \tilde{\bm \chi}\,.
\ee

\subsubsection*{Semi-classical Einstein's equations}

We are now ready to study the implications of the corrected dictionaries (\ref{deltaSIW}) and (\ref{deltaEIW}). First, specializing to a slice $\Sigma_A$ at constant$-t$ and making use of the first law of entanglement entropy we arrive at
\be
\delta S_A^{\text{grav}}-\delta E_A^{\text{grav}}=\int_{\gamma_A}\tilde{\bm \chi}-\int_A \tilde{\bm \chi}+\int_{\Sigma_A}\!\! \xi^t \langle T^{\text{bulk}}_{00}(x)\rangle {\bm \epsilon}^t=0\,.
\ee
Using Stoke's theorem and using the fact that in this slice $d\tilde{\bm \chi}=-2\xi^{t}\delta E^{g}_{00}{\bm \epsilon}^t$, with $E^{g}_{\mu\nu}$ given in (\ref{linearizedEab}), we can now rewrite the surface terms as a volume term,
\be
\delta S_A^{\text{grav}}-\delta E_A^{\text{grav}}=-2\int_{\Sigma_A}\!\! \xi^t \left(\delta E_{00}-\frac{1}{2}\langle T^{\text{bulk}}_{00}(x)\rangle\right) {\bm \epsilon}^t=0\,.
\ee
Considering all possible balls $A$ in $\Sigma_A$
\be
\delta E_{00}=\frac{1}{2}\langle T^{\text{bulk}}_{00}(x)\rangle
\ee
at each bulk point \cite{Swingle:2014uza}, i.e., the 00 component of the linearized semi-classical Einstein's equations universally coupled to matter! We can further repeat this analysis by specializing to various boosted frames, labeled by a velocity vector $u^\mu$. As a result, we obtain all other components of the semi-classical Einstein's equations,
\be
u^\mu u^\nu\left(\delta E_{\mu\nu}-\frac{1}{2}\langle T^{\text{bulk}}_{\mu\nu}(x)\rangle\right)=0\,.
\ee
Remarkably, the above analysis implies that if we identify the perturbed thread configuration as in \cite{Agon:2020mvu}, given in equation (\ref{w-chi}),
the corrected dictionaries automatically imply that $\delta \bm w$ is a consistent solution to the max flow problem at the desired order in $G_N$. Notice that this proposal makes complete use of bulk locality, since the semi-classical Einstein's equations are explicitly encoded in the condition
\be
d(\delta \bm w)=-8G_N\xi^{\mu}\delta E^{g}_{\mu\nu}{\bm \epsilon}^\nu=-4G_N\xi^{\mu}\langle T^{\text{bulk}}_{\mu\nu}(x)\rangle{\bm \epsilon}^\nu=-4G_N s(x)\,,
\ee
while at the same time, giving rise to the known entanglement density for bulk regions with local modular Hamiltonians (\ref{bulkMH}), i.e., equation (\ref{density:localMH}).

%

\section{Discussion\label{sec:Disc}}

In this paper, we have shown that the leading quantum corrections to holographic entanglement entropy can be interpreted in terms of a generalized flow $v$, where
\be\label{QBTresult}
S_A=\frac{1}{4G_N}\, \max_{v\in \mathcal{F}}\int_{A} v\,,\qquad {\cal F}\equiv\{v\, \vert\, \nabla\cdot v=-4 G_Ns(x),\, |v|\leq 1\}\,,
\ee
and
\be
\int_{\Sigma_A} \!s(x) = S_{\text{bulk}}[\Sigma_A]\,.
\ee
To fully specify the program, then, we are required to provide an entanglement density $s(x)$ for the bulk homology region $\Sigma_A$. A couple of comments are in order. First, we note that there is no way to provide a local density $s(x)$ that is valid for \emph{any} bulk subregion. Thus, different regions $A_i$ with their associated bulk homology regions will generally have different flow programs. In some special cases, though, it is possible to come up with a reasonable density $s(x)$ that is valid for a number of regions, and then define a combined program. This is possible in two cases i) when the regions are non-overlapping or ii) when the regions are strictly nested. One important point that needs emphasis is that this prescription is meant to be used to interpret the quantum corrections, but it should not be though of as a computational tool in the same sense as the standard prescription (\ref{BTpres}). The fact that we need to specify a density $s(x)$ to fully specify the program means, among other things, that we are already starting with the basic ingredients needed to compute the entropy, that is, the minimal area surface and the bulk entropy. Conversely, this also means that we cannot allow the flow maximization to backreact on the density $s(x)$, and hence on the homology region $\Sigma_A$. Translating back to the original minimization problem, this implies that we must restrict ourselves to the leading order quantum corrections only, that is, to order $\mathcal{O}(G_N^0)$. As mentioned in the introduction, the FLM and QES formulas can differ at this order in situations close to a phase transition.  However, for the derivation of the above program we have used the QES formula as a starting point, and thus our prescription should be valid and useful to diagnose quantum phase transitions. Again, we emphasize that our proof only holds at the given order. It is only in this case the problem can be cast as a convex program and hence we can use strong duality. We could speculate that a version similar to (\ref{QBTresult}) could hold true for the full QES prescription. However, it is difficult to imagine that this can be done in full generality, because it is not possible to come up with a density that is valid for \emph{any} region and holds universally for any theory/state. Even so, in very special cases this may be feasible. We will comment on this possibility in the outlook.

An important aspect of our prescription is its physical interpretation. In section \ref{sec:interp} we argued that a discrete version of our formula can be equivalently thought of in terms of a set of classical and quantum Planck-thickness bit threads. Roughly speaking, we consider classical threads as those which connect boundary points with other boundary points, and quantum bit threads as those which connect either boundary and bulk points or bulk and bulk points. Thus, quantum threads are those which are sourced by the bulk entropy density. This separation turns out not to be unique, though, as one can always break a classical thread into two quantum threads or viceversa by the expense of having a pair of oppositely charged bulk entropy density `units' at the breaking point. This process is reminiscent to the breaking of a gravitational Wilson line, as explained in \cite{Harlow:2015lma}. It would be very interesting to try to interpret bit threads in terms of the latter, which we leave for a future exploration. A further observation is that one can also interpret a pair of oppositely charged units of bulk entropy density as representing a bulk Bell pair. In this picture, then, one can view the extra flux through $A$ arising from the presence of quantum threads on both sides of the minimal surface as being sourced by the presence of bulk Bell pairs. In summary, then, one can interpret our formula as giving the total number of distilled boundary Bell pairs between $A$ and $\bar{A}$; some of these are represented geometrically via continuous threads that directly connect the regions $A$ and $\bar{A}$ through the bulk, while others connecting them via subtle bridges created by the presence of the bulk Bell pairs (via `ER=EPR' \cite{Maldacena:2013xja}). This discrete interpretation of our formula was advocated throughout section \ref{sec:GeneralProps} to prove some non-trivial properties of our max flow program, including quantum generalizations of nesting and the max multiflow theorem of classical threads. Later in the same section we used the nesting property show some basic properties that the boundary entropy must satisfy, such as subadditivity and strong subadditivity and thus providing significant evidence for the validity of our program. We concluded this section by using the quantum version of the max multiflow theorem to investigate the necessary conditions to obtain boundary monogamy. We obtained that the geometrization of the bulk entropies is an essential ingredient for this inequality to hold. This geometrization implies bulk monogamy, however, the contrary is not true. Thus, from our perspective, bulk monogamy is required but perhaps not sufficient for boundary monogamy to hold beyond leading order in $1/N$.\footnote{Very recently we became aware of \cite{Akers:2021lms} which claims that bulk monogamy is indeed enough to ensure boundary monogamy.}

We finished our work in section \ref{sec:IyerWald} by applying our prescription to the case of perturbative, semi-classical bulk states. These perturbative states can encode non-trivial dynamics. However, one work around using the fully covariant formulation of bit threads by specializing to the same Cauchy slice $\Sigma$ used to solve the max flow problem in the vacuum case (i.e., empty AdS). As explained in \cite{Agon:2020mvu} this is consistent, and can be justified if one starts from the maximin formulation of the HRT formula. With this working assumption, then, we were able to show that the Iyer-Wald construction presented in \cite{Agon:2020mvu} naturally encodes the leading quantum corrections when applied to bulk regions with local modular Hamiltonians, given the class of bulk states in consideration. This means that the Iyer-Wald formalism provides us with a canonical choice for the perturbed thread configuration that solves the quantum max flow problem. Furthermore, this particular solution makes bulk locality manifest, a property that was essential in \cite{Agon:2020mvu} for the problem of metric reconstruction. Combining this result with previous work by Swingle and Van Raamsdonk \cite{Swingle:2014uza}, we were able to show that this special solution implies that the \emph{semi-classical} Einstein's equations must hold for any consistent perturbative bulk quantum state. Semi-classical gravity is then seen to arise from entanglement considerations in the dual CFT and consistency of our quantum max flow prescription.

We conclude with a list of some open questions that we think are worth exploring:
\begin{itemize}
  \item \emph{Fully covariant generalization:} In this paper we have mostly focussed on static states, or small perturbations of static states. In these situations the standard bit thread prescription is valid (in the latter case invoking a restricted version of the maximin formulation of HRT \cite{Agon:2020mvu}). In fully dynamical situations, however, one needs to upgrade the recipe and use the full covariant formula instead. We note that a covariant proposal of bit threads will appear soon in \cite{Headrick:toappear}. It would be very interesting to upgrade this prescription to include quantum corrections and try to apply it to situations with a dynamical black hole in the bulk. For collapsing solutions, this would  shed light on the physical interpretation of the \emph{entanglement tsunami} proposal for entanglement propagation \cite{Liu:2013iza,Liu:2013qca} and its breakdown \cite{Kundu:2016cgh,Lokhande:2017jik}. Conversely, for evaporating solutions, this could clarify the interpretation of the so-called quantum extremal islands which were recently proposed in the context of the information loss problem \cite{Penington:2019npb,Almheiri:2019psf,Almheiri:2019hni,Almheiri:2019qdq}.
 \item \emph{Thermodynamic limit:} Finite temperature states or \emph{static} black hole geometries are also interesting on its own right. In the thermodynamic limit, i.e., for large enough regions the entanglement entropy should converge to thermal entropy and become extensive. In this limit, one could expect the bulk entropy to be additive and thus one would actually have an entanglement density (thermodynamic density) such that its integral on any (large) bulk subregion would reproduce the leading term of the bulk entanglement entropy. This would provide a simple scenario in which our prescription would reproduce the full QES answer.  All the properties such as subadditivity and strong subadditivity would immediately follow as they would automatically be saturated. This framework could be useful to interpret results concerning quantum corrections to black hole entropy.
  \item \emph{Models of double holography:} Double holographic setups provide us with an interesting tool to understand the possible flow program dual to the full QES prescription more generally. This is because in these setups the full quantum corrections are completely geometrized in a higher dimensional space. Given so, one could imagine using the standard bit thread prescription in the higher dimensional space and then project the resulting flows back onto the lower dimensional space. This projection will induce a source term for the lower dimensional vector field, which will generally depend on the particular outcome of the original optimization problem. As a whole, then, one could think of this higher dimensional program as way of deriving an \emph{density functional}, encoding all possible densities associated with all possible bulk regions. Evidently, this is infinitely more information of the bulk state than the minimal input we need to provide for the leading order prescription. Alternatively, we could try to completely translate the higher-dimensional program into quantities defined in the lower dimensional space. This will likely involve a non-linear mapping between variables and could shed light on the correct implementation of the full QES prescription in a more general setting.
  \item \emph{Higher order inequalities:} In this work we showed that the subadditivity and strong subadditivity inequalities for boundary entropies follow from the nesting property of quantum bit threads and the assumption that the bulk entropies satisfy the same inequalities. This is justified, though, since these inequalities must be obeyed in any consistent quantum theory. A natural question that we can ask is: what happens with higher order inequalities of the holographic entropy cone \cite{Bao:2015bfa,Cuenca:2019uzx}? Our preliminary results suggest that the monogamy of mutual information can be derived using our quantum generalization of the max multiflow theorem, as was done in the classical case \cite{Cui:2018dyq}, although requiring a `geometrization condition' for the bulk entropies. This is a \emph{sufficient} condition in our proof, however, it certainly seems stronger than bulk MMI (see, however \cite{Akers:2021lms}). There are reasons to further investigate this question, however, in order to clarify the consistency conditions of double holographic scenarios. It would be interesting to try to relax our proof of boundary MMI, and ask what are the \emph{minimal} constraints we need to impose for the bulk entropies. Similarly, it would be interesting to investigate the fate of higher order inequalities of the holographic entropy cone.
  \item \emph{Other information-theoretic observables:} Another interesting question we can ask is if there are other information-theoretic quantities whose quantum corrections admit a quantum bit thread interpretation. We give here two examples. First, the entanglement wedge-cross section. There are various CFT quantities that have been linked to this bulk observable, including entanglement of purification \cite{Takayanagi:2017knl,Nguyen:2017yqw}, reflected entropy \cite{Dutta:2019gen}, logarithmic negativity \cite{Kudler-Flam:2018qjo}, odd entropy \cite{Tamaoka:2018ned} and balanced partial entanglement \cite{Wen:2021qgx}. Understanding their quantum corrections could help determine if there is still a connection beyond leading order in $1/N$ and, hence, discriminate between these proposals. And second, the so-called differential entropy \cite{Balasubramanian:2013lsa}. A flow prescription for this quantity would naively involve some optimization process subject to having some charges along the bulk curve in consideration, which is reminiscent of our quantum flow program. It would be interesting to see if quantum threads could shed light on its boundary interpretation.
  \item \emph{Non-linear Einstein's equations:} Finally, we could ask if it is possible to go to higher orders in the perturbation and check if the full semi-classical Einstein's equations arise consistently from our canonical flows. We note that, from the point of view of the RT prescription, the first non-linear correction was successfully derived in \cite{Faulkner:2017tkh}, although, restricting to classical perturbations. In \cite{Swingle:2014uza}, the authors argued that
      if one assumes that the local dynamical equations that are derived at the linearized level extend to some local non-linear equations, and demand self-consistency of these equations with conservation of energy and momentum, then it is almost immediate that these equations must be the full Einstein's equations. It would be very interesting to explicitly check this idea, at least at the first non-linear order, which could be done by extending the work of \cite{Faulkner:2017tkh} to the case of semi-classical bulk states (the ones we considered in this work) and then constructing the non-linear version of our canonical flows.
\end{itemize}
We hope to come back to some of these points in the near future.

\paragraph{Note added:} While we were at the final stages of writing up our paper we became aware of the work of \cite{Rolph:2021hgz} whose results partially overlap with ours. The two papers are being submitted simultaneously.

\section*{Acknowledgements}

It is a pleasure to thank Ning Bao, Elena C\'aceres, Eoin Colg\'ain, Jan de Boer, Brandon DiNunno, Matthew Headrick, Andrea Russo, Andrew Svesko, Zach Weller-Davies and Qiang Wen for useful discussions and comments on the manuscript and to Harsha Hampapura for collaboration during the early stages of the project. This material is based upon work supported by the Simons Foundation through \emph{It from Qubit: Simons Collaboration on Quantum Fields, Gravity, and Information}.

\bibliographystyle{ucsd}
\bibliography{refs-QBT}

\end{document}